\documentclass[twocolumn]{aastex701}
\usepackage{graphicx} 
\usepackage{amsmath}
\usepackage{amssymb}

\newcommand{\vv}{{\tablenotemark{\footnotesize{a}}}}
\newcommand{\vturb}{\ensuremath{{\xi}}}
\newcommand{\kmsec}{\mbox{km~s$^{\rm -1}$}}

\newcommand{\eps}[1]{\ensuremath{\log\epsilon\,(\mathrm{#1})}}
\newcommand{\abund}[2]{\ensuremath{[\mathrm{#1}/\mathrm{#2}]}}
\newcommand{\cfe}{\abund{C}{Fe}}

\newcommand{\xfe}[1]{\abund{#1}{Fe}}

\newcommand{\metal}{\abund{Fe}{H}}

\newcommand{\teff}{\ensuremath{T_\mathrm{eff}}}
\newcommand{\logg}{\ensuremath{\log\,g}}

\newcommand{\A}[1]{\ensuremath{A(\mathrm{#1})}}

\newcommand{\K}{\ensuremath{\mathrm{K}}}

\newcommand{\bdp}{\object{\mbox{BD$+$44$^{\circ}$493}}}

\begin{document}

\title{Near-Infrared Observations of \bdp\, with IGRINS-2}

\author[0000-0003-4479-1265]{Vinicius M.\ Placco}
\affiliation{NSF NOIRLab, Tucson, AZ 85719, USA}
\email{vinicius.placco@noirlab.edu}

\author[0000-0002-9123-0068]{William D. Vacca}
\affiliation{NSF NOIRLab, Tucson, AZ 85719, USA}
\email{bill.vacca@noirlab.edu}

\date{\today}

\begin{abstract}

We present high-resolution ($\mathcal{R}\sim45,000$), high signal-to-noise (S/N$>200$) H and K-band spectra of \bdp\, obtained with the newly commissioned IGRINS-2 instrument on the Gemini North Telescope. \bdp\, is a well-known carbon-enhanced (CEMP; \cfe$=+1.40$) ultra metal-poor (UMP; \metal$\sim-4.0$) star that has been extensively studied with high-resolution spectroscopy from the ultraviolet to the infrared. From the IGRINS-2 data, we derive new estimates of the abundances of C, Mg, Si, and Ca by measuring the equivalent widths of several atomic absorption features and comparing the values and line profiles with synthetic models. We measure, for the first time, a weak atomic carbon absorption feature (\ion{C}{1}) in a UMP star in the H-band. Our estimated abundance agrees well with other values derived from CH molecules in the optical. The abundance values for the other elements are also in good agreement with previous estimates derived from ultraviolet, optical, and near-infrared observations, confirming the speculations that \bdp\, is a bona fide second-generation star. With these data, we demonstrate that IGRINS-2 is a powerful new resource for studying and characterizing low-metallicity stars in the Milky Way and beyond.

\end{abstract}


\section{Introduction}

Metal-poor (MP) stars \citep[\metal ~ $\leq-1.0$\footnote{\abund{A}{B} = $\log(N_A/{}N_B)_{\star} - \log(N_A/{}N_B) _{\odot}$, where $N$ is the atom number density of an element in the star ($\star$) and the Sun ($\odot$).};][]{frebel2015} are the surviving witnesses of the Milky Way's chaotic past, retaining in their atmosphere the chemical fingerprints of the evolution of the first stars (Population III) to be formed in the Universe after the Big Bang \citep{abel2002, bromm2004, bromm2009}. Through chemo-dynamical analysis of these old, low-mass, chemically peculiar stars, it is possible not only to reconstruct the assembly history of the Galaxy, but also to investigate the stellar populations and the nucleosynthetic processes that have operated since the onset of star formation in the early Universe \citep{bonifacio2025}.

The determination of the detailed chemical inventory of such unique objects is possible only with high-resolution ($\mathcal{R}=\lambda/\Delta\lambda\geq20,000$) spectroscopy. Most of the targeted efforts to follow up metal-poor star candidates identified from large-scale surveys (e.g., SDSS - \citealt{york2000}, Hamburg/ESO Survey - \citealt{christlieb2008}, Pristine - \citealt{starkenburg2017}, Sky Mapper - \citealt{wolf2018}, S-PLUS - \citealt{mendesdeoliveira2019}) have concentrated on the optical wavelength regime ($\lambda\sim3,200-9,000$~\AA), where hundreds of absorption features of interest can be available for each star, covering lithium to uranium abundances \citep{roederer2018}. A few notable examples of high-resolution spectroscopic follow-up efforts of low-metallicity stars include the following programs: ``Extremely Metal-Poor Stars'' \citep{norris1996}, ``First Stars'' \citep{cayrel2004}, ``The Most Metal-Poor Stars'' \citep{norris2013}, ``A Search for Stars of Very Low Metal Abundance'' \citep{roederer2014}, and ``R-Process Alliance'' \citep{hansen2018}.

In the ultraviolet regime ($\lambda\sim1,150-3,200$\AA), extensive follow-up of metal-poor stars has been done with the COS and STIS spectrographs on the Hubble Space Telescope \citep{sneden1998,cowan2002,roederer2012b,siqueira2013,placco2015b,ernandes2023,hansen2025,roederer2025}. Besides additional absorption features for elements commonly measured in the optical (e.g. carbon, oxygen, zinc) there are also several important species with transitions available only in the ultraviolet, such as boron, phosphorus, germanium, selenium, tin, among others \citep{cunha2000,roederer2023,spite2025}.

In the near-infrared (NIR) regime, a transformative project is APOGEE \citep[Apache Point Observatory Galactic Evolution Experiment;][]{majewski2017}, which has been collecting $\mathcal{R}\sim22,500$ spectra in the $H$ band ($\lambda\sim1.5-1.7\mu\rm{m}$) since 2011, totaling over 150,000 stars by the end of APOGEE-1. Although this project primarily targeted the inner Galaxy, it collected data on several low-metallicity stars, leading to subsequent abundance determinations \citep[e.g.,][]{hasselquist2016,hayes2018,trincado2019,razera2022,barbuy2024}. For more focused, single-star studies, the IGRINS \citep[Immersion GRating INfrared Spectrometer;][]{yuk2010,mace2016} instrument has provided simultaneous coverage of the $H$ and $K$ bands ($\lambda\sim1.49-2.46\mu\rm{m}$) at $\mathcal{R}\sim45,000$. Since commissioning in July 2014, IGRINS has been offered on the 2.7-meter Harlan J. Smith Telescope at McDonald Observatory, the 4.3-meter Discovery Channel Telescope at Lowell Observatory, and the 8.1-meter Gemini South telescope (until early 2024).

For metal-poor stars, the NIR window provides access to atomic C lines, CO vibrational bandheads, OH lines, and CN features that independently constrain carbon, oxygen, and nitrogen abundances, while offering reduced sensitivity to continuum placement uncertainties that affect blue optical spectra at low metallicity. In addition, NIR spectra provide key diagnostic lines for Mg, Si, Ca, and Ti, enabling consistency checks with optical abundance determinations, as well as species without suitable transitions in the optical, such as phosphorus \citep{afsar2016,mura2020,nandakumar2022}. Recently, the IGRINS-2 instrument (with the same capabilities as IGRINS) was commissioned at the Gemini North telescope. This adds a much-needed capability on a large open-access telescope in the Northern Hemisphere, complementing the IRD \citep[InfraRed Doppler;][]{tamura2012} instrument at the Subaru Telescope, which covers the $Y$, $J$, and $H$ bands at $\mathcal{R}\sim70,000$. \footnote{We note that, among other instruments, the iSHELL spectrograph at the IRTF covers the 1-5 $\mu$m range at a resolution of $\mathcal{R} \sim 80,000$ \citep{rayner2022}. However, multiple settings and integrations are required to cover the $H$ and $K$ bands with this instrument, and the faintness of potential targets is limited by the telescope's 3-meter aperture.}

\bdp\, is a bright ($V \sim 9.11$), old (12.1-13.2~Gyr), low mass ($\sim0.83M_\odot$), carbon-enhanced \citep[CEMP: \cfe$\geq0.7$;][]{aoki2007}, ultra metal-poor (UMP; \metal$\sim -4.0$) sub-giant star in a disk-like orbit, at a distance of about 200 parsecs. Its high carbon abundance, coupled with a low abundance of neutron capture elements, places \bdp\ in the CEMP-no Group~III class according to \citet{yoon2016}. Due to its brightness and peculiar chemistry, \bdp\, has been extensively studied in the literature, with spectroscopic studies spanning the ultraviolet \citep{placco2014b,roederer2016}, optical \citep{ito2009,ito2013,placco2024}, and infrared \citep{takeda2013,aoki2025} wavelength regimes. A comprehensive summary of previous work on the properties of \bdp\, has recently been provided by \citet{placco2024}. 

In this paper, we present new $H$ and $K$ band spectra of \bdp\, obtained with the IGRINS-2 spectrograph at the Gemini North telescope. This chemically unusual star is a perfect testbed for assessing the capabilities of the IGRINS-2 spectrograph in the high S/N regime, primarily because of its temperature (\teff=5351~K) and metallicity, for which only a few weak absorption features are expected in the near-infrared. We obtained high S/N ($\sim 300$) data with a total on-source exposure time of only 12.5 min. The primary purpose is to demonstrate the capabilities of IGRINS-2 for abundance determinations, particularly for low-metallicity objects. The paper is organized as follows: Section~\ref{sec:obs} describes the \bdp\, IGRINS-2 observations and data reduction, followed by the line measurements in Section~\ref{sec:line} and chemical abundance determinations in Section~\ref{sec:abund}. A brief discussion and perspectives for future work are given in Sections~\ref{sec:disc} and \ref{sec:conc}, respectively.

\begin{figure*}[ht!]
\includegraphics[height=0.6\textheight,angle=180]{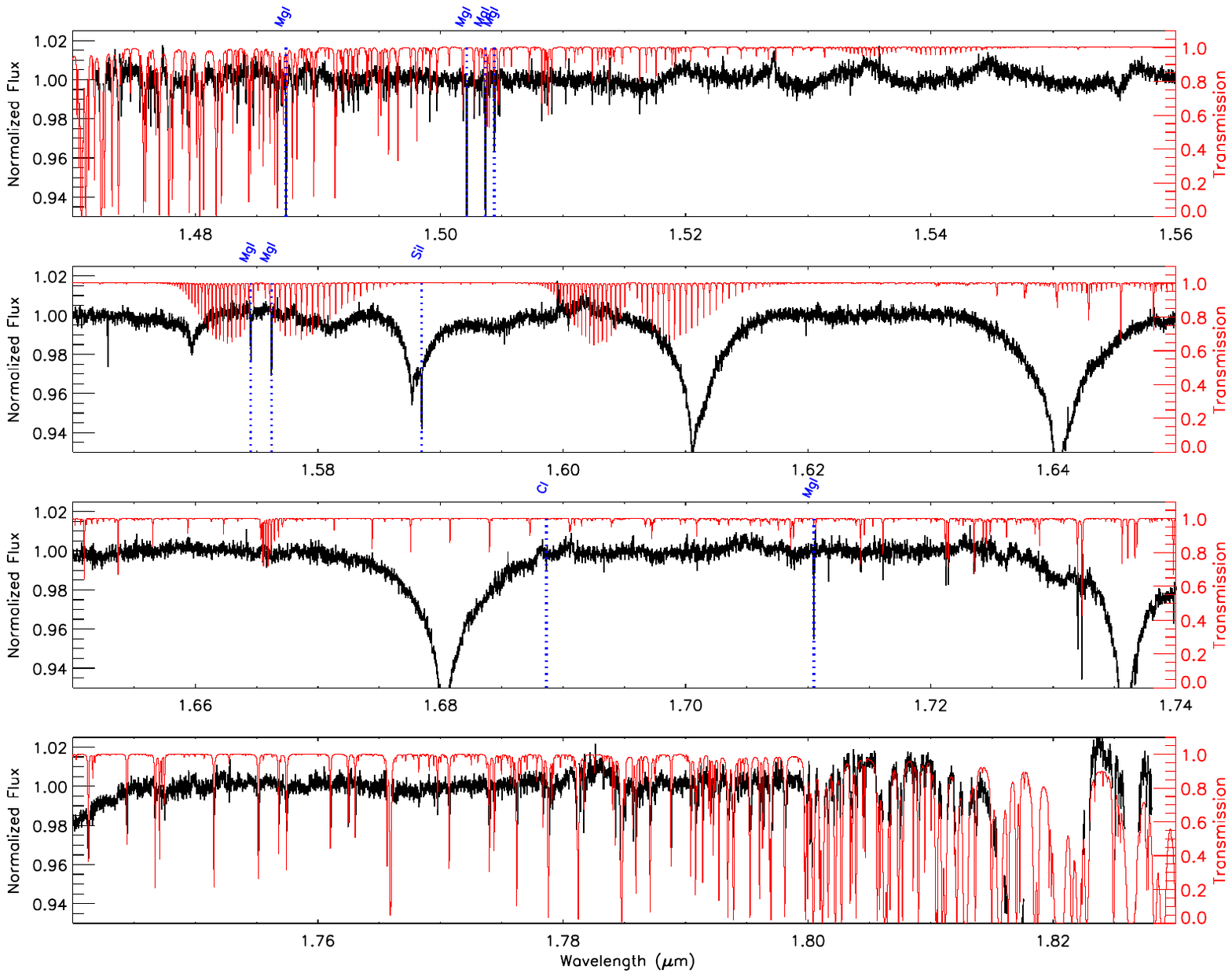}
\caption{Telluric-corrected H band spectrum of \bdp. Wavelengths are in vacuum, and no shift to the rest frame of \bdp\, has been applied. A nominal atmospheric spectrum for Gemini North is overplotted in red. Metal absorption lines are indicated by vertical blue dashed lines, and their corresponding IDs are provided at the top. The strong absorption lines with the sharp cores at 1.588, 1.611, 1.641, 1.681, 1.736 $\mu$m are the H Brackett lines in the spectrum of \bdp. The small emission bumps between 1.52 and 1.56 $\mu$m and the broad shallow red wings on the H lines are residuals from the telluric correction procedure and are due to the H Brackett series in the spectrum of the telluric standard. 
}
\label{fig:Hplot_panel}
\end{figure*}

\begin{figure*}[ht!]
\includegraphics[height=0.6\textheight,angle=180]{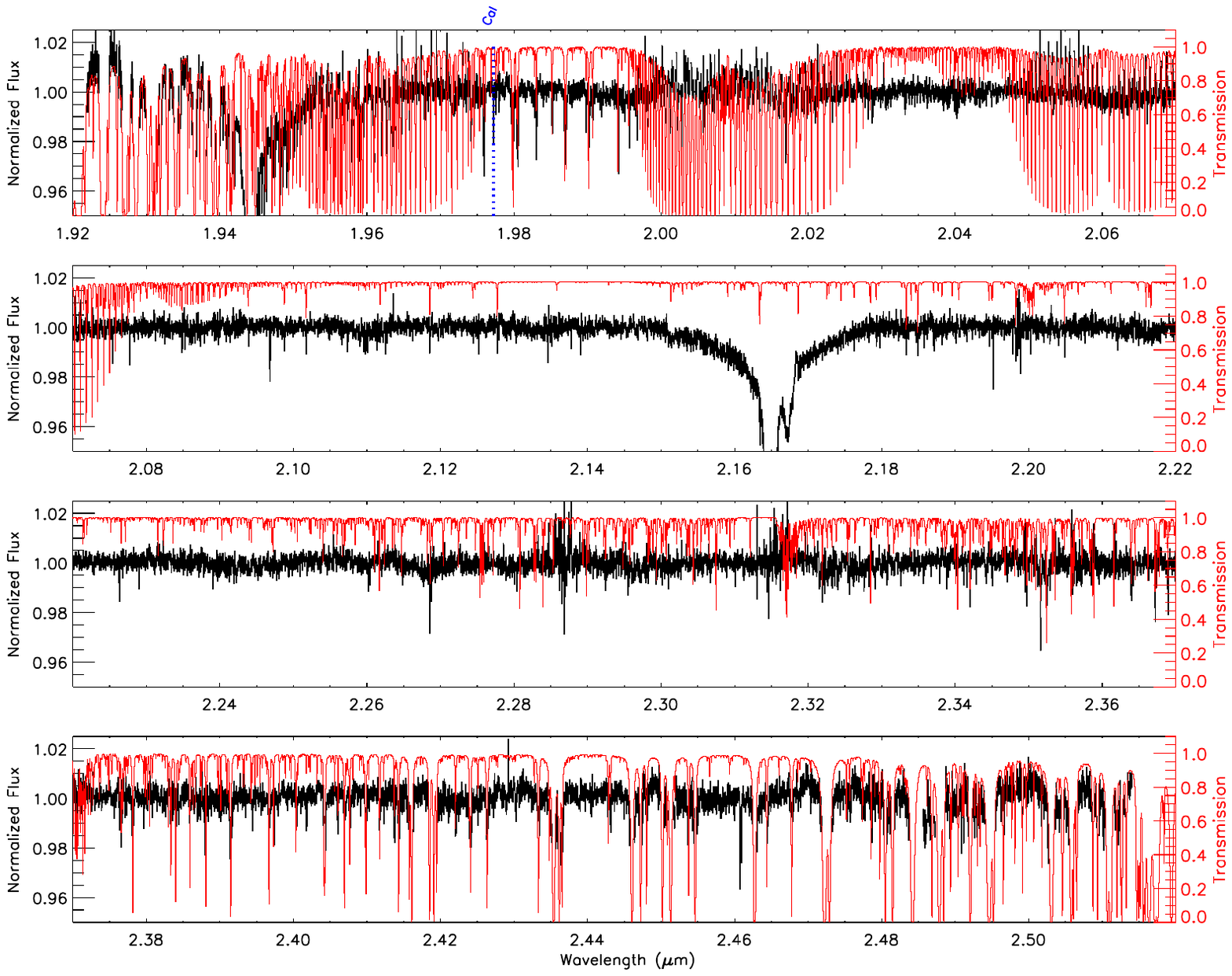}
\caption{Telluric-corrected K band spectrum of \bdp. Wavelengths are in vacuum, and no shift to the rest frame of \bdp\, has been applied. A nominal atmospheric spectrum for Gemini North is overplotted in red. Metal absorption lines are indicated by vertical blue dashed lines, and their corresponding IDs are given at the top. The deep absorption features at 1.945 and 2.166 $\mu$m are the Br lines in the spectrum of \bdp. The shallow dip seen in the red wing of the Br $\gamma$ line is a residual from the telluric correction procedure and is due to Br $\gamma$ absorption in the spectrum of the telluric standard. 
}
\label{fig:Kplot_panel}
\end{figure*}

\begin{deluxetable}{lrrrrrrrrrrrrrrrrccc}
\tabletypesize{\tiny}
\tabletypesize{\footnotesize}
\tablewidth{0pc}
\tablecaption{\label{tab:syn} Atomic Data and Derived Abundances}
\tablehead{
\colhead{Ion}&
\colhead{$\lambda_0$}&
\colhead{$\lambda_0$}&
\colhead{$\lambda_{\rm obs}$}&
\colhead{$\chi$} &
\colhead{$\log\,gf$}&
\colhead{EW}&
\colhead{$\sigma_{\rm EW}$}&
\colhead{$\log\epsilon$\,(X)}&
\colhead{$\log\epsilon$\,(X)}&
\colhead{$\Delta$NLTE}\\
\colhead{}&
\colhead{({\AA\, - air})}&
\colhead{({\AA\, - vac})}&
\colhead{({\AA\, - vac})}&
\colhead{(eV)} &
\colhead{}&
\colhead{(m{\AA})}&
\colhead{(m{\AA})}&
\colhead{EW}&
\colhead{syn}&
\colhead{}}
\startdata
\ion{C}{1}  & 16890.38 & 16894.87 & 16886.39 &  9.003 &    0.570 &  4.23 & 0.42 &  5.59 & 5.74 & \nodata \\ 
\ion{Mg}{1} & 14877.53 & 14881.62 & 14874.15 &  5.946 &    0.690 & 51.23 & 0.78 &  4.83 & 4.34 &    0.31 \\
\ion{Mg}{1} & 15024.99 & 15029.01 & 15021.47 &  5.108 &    0.330 & 53.29 & 0.86 &  4.35 & 4.31 &    0.09 \\
\ion{Mg}{1} & 15040.25 & 15044.25 & 15036.70 &  5.108 &    0.110 & 35.27 & 0.76 &  4.34 & 4.34 &    0.09 \\
\ion{Mg}{1} & 15047.70 & 15051.68 & 15044.13 &  5.108 & $-$0.360 & 23.19 & 0.75 &  4.60 & 4.56 &    0.09 \\
\ion{Mg}{1} & 15748.99 & 15753.17 & 15745.26 &  5.932 &    0.140 & 15.17 & 0.75 &  4.63 & 4.46 & \nodata \\
\ion{Mg}{1} & 15765.84 & 15770.08 & 15762.16 &  5.928 &    0.380 & 19.57 & 0.76 &  4.51 & 4.35 &    0.27 \\
\ion{Mg}{1} & 17108.63 & 17113.23 & 17104.65 &  5.394 &    0.064 & 25.41 & 0.87 &  4.37 & 4.29 &    0.06 \\
\ion{Si}{1} & 15888.39 & 15892.67 & 15884.69 &  5.082 & $-$0.002 & 17.48 & 0.61 &  3.98 & 3.99 & $-$0.02 \\
\ion{Ca}{1} & 19776.77 & 19782.04 & 19772.11 &  1.897 & $-$0.830 & 17.03 & 1.02 &  3.08 & 2.88 & \nodata \\
\enddata
\tablerefs{
NLTE corrections -- 
\ion{Mg}{1}: \citet{bergemann2015};
\ion{Si}{1}: \citet{bergemann2013}.
}
\end{deluxetable}
\section{Observations and Data Reduction}
\label{sec:obs}

The observations were carried out in queue mode on two nights (04 and 06 Nov 2025 UT) using the IGRINS-2 instrument \citep{choi2025, oh2024, lee2022} at Gemini North in Band 4 (poor weather) conditions (Program ID: GN-2025B-Q-415\footnote{\href{https://archive.gemini.edu/searchform/GN-2025B-Q-415}{https://archive.gemini.edu/searchform/GN-2025B-Q-415}}). IGRINS-2 is a cross-dispersed high-resolution ($\mathcal{R} \sim 45,000$) spectrograph covering the $H$ ($1.44-1.84~ \mu$m) and $K$ ($1.89 - 2.55~ \mu$m) bands. The slit has a fixed width of $0.33\arcsec$ and a length of $5\arcsec$ and was aligned with the parallactic angle during the observations. Data were obtained using the standard ABBA nodding sequence with the A and B positions separated by 2.5$\arcsec$. Four frames were obtained for \bdp, each with an exposure time of 188 s, in both the $H$ and $K$ bands. The airmass of the observations was 1.105 on 04 Nov and 1.135 on 06 Nov. The weather conditions were at the 70th percentile for image quality (\texttt{IQ70}) and the 80th percentile for cloud cover (\texttt{CC80}), with no restrictions on sky brightness or water vapor. Four frames were taken for the telluric standard (HD 21620, A0Vn\footnote{Although HD 21620 has a spectral type of A0Vn, we realized after the fact that this was a poor choice for a telluric standard, since the star is also an eclipsing binary with a circumstellar gas and dust debris disk, and perhaps even infalling exocomets \citep{welsh2013}. Fortunately, we believe these features had minimal impact on the observed spectrum of \bdp\, except for weak absorption artifacts in the red wings of some of the H line profiles; see Figures~\ref{fig:Hplot_panel} and \ref{fig:Kplot_panel}.}), at a similar airmass ($\Delta Z = -0.062$ on 04 Nov and $-0.016$ on 06 Nov), each with an exposure time of 80 s, immediately after the observations of \bdp. 

The data were reduced using version 3.2 of the IGRINS-2 data-reduction pipeline \citep{sim2014,sawczynec2025,kaplan2024}. The data reduction software performs the typical reduction steps necessary to process NIR data: combining the frames for each position, subtracting the combined B frames from the combined A frames to remove sky emission lines, applying a flat field to the result, rectifying the individual orders, determining a wavelength solution, and extracting the spectrum for each order. Additional details on the reduction steps are described by \citet{sim2014} and \citet{sawczynec2025}.

The reduced spectra were corrected for telluric absorption and flux-calibrated using the observations of the A0V star (HD 21620) and a modified version of the \texttt{xtellcor} software package \citep{vacca2003}. This software assumes that the observed spectra of A0V stars can be represented as scaled versions of the Vega spectrum, which is modeled using the theoretical spectrum computed by \citet{kurucz1992}. The theoretical spectrum was broadened to account for the known rotational velocity of HD~21620\footnote{In an attempt to recover symmetric intrinsic H line profiles in the telluric-corrected spectrum of \bdp\, we also increased the rotational broadening applied to the Vega model beyond the known value for HD~21620. However, as shown in Figures~\ref{fig:Hplot_panel} and \ref{fig:Kplot_panel}, we were only partially successful, as no single value yielded consistent results simultaneously for all of the H lines in the spectra of \bdp\ in both the $H$ and $K$ bands.}, shifted to the known radial velocity, smoothed to the resolution delivered by IGRINS-2, interpolated to the observed sampling, and then scaled to match the observed magnitudes of HD 21620. Dividing the observed spectrum of HD~21620 by the smoothed and interpolated model yielded the telluric correction model. This was then applied to the \bdp\, spectrum. 

Small shifts were applied to each order of the telluric spectrum to account for wavelength shifts between the target and the standard. The shifts were determined separately for each order, but were found to be fairly consistent across the orders. For the data acquired on 04 Nov, the shifts were found to be $0.20 \pm 0.07$ pixels in the $H$ band and $0.33 \pm 0.04 $ pixels in the $K$ band. The shifts were considerably smaller for the data acquired on 06 Nov, with mean values of $-0.02 \pm 0.02$ pixels for the $H$ band and $ 0.02 \pm 0.01$ pixels for the $K$ band. Although small, these adjustments to the wavelength scale can substantially reduce residual artifacts in the target spectrum. Nevertheless, even after applying these shifts, the spectra from 04 Nov exhibited prominent telluric residuals. The larger shifts measured for the data obtained on 04 Nov reflect the significantly poorer quality and signal-to-noise ratio (S/N) of these spectra compared to those obtained on 06 Nov, despite the higher airmass at which the latter were obtained. For this reason, we chose not to analyze the spectra from 04 Nov further, and we consider only the data obtained on 06 Nov for the remainder of this paper. 

As shown in Figures~\ref{fig:Hplot_panel} and \ref{fig:Kplot_panel}, the telluric correction procedure, although not perfect, successfully recovered much of the target spectrum on this night, even in regions with numerous narrow telluric features (e.g., near $2.06 \mu$m). The spectrum in each order was `cleaned' by removing pixels with S/N ratios less than 100\footnote{Applying an SNR cut of 100 preserves $\sim82\%$ of the spectrum in both the H and K bands. Most of the pixels removed by this cut are in the orders at the very tops and bottoms of the arrays where the instrumental throughput is relatively low and there is only partial coverage of the spectrum.}, the ends of the individual orders were trimmed, and the orders were then spliced together to generate a single telluric-corrected, flux-calibrated spectrum of the target in the $H$ and $K$ bands. The final S/N ratios of the reduced spectra vary considerably across each order, with the highest values found in the centers of each order, but are typically between $\sim 200$ and $\sim 400$ for almost all orders in both bands.  

\section{Line Measurements}
\label{sec:line}

The $H$ and $K$-band spectra for \bdp\, were normalized by fitting spline functions with sigma-clipping to each order using the {\texttt{Spectroscopy Made Harder}} \citep[\texttt{SMHr};][]{casey2014}\footnote{\href{https://github.com/andycasey/smhr}{https://github.com/andycasey/smhr}} software. The normalized spectra are presented in Figures \ref{fig:Hplot_panel} and \ref{fig:Kplot_panel}. 
Pixels corresponding to possible absorption lines were identified in the normalized spectra both visually and automatically. In the latter case, we measured the noise in the normalized spectra over a span of 100 pixels in the vicinity of any possible line and applied a S/N criterion for the pixel value at the putative line center of 5 (see Appendix~\ref{sec:app} for details), taking into account a telluric absorption spectrum generated by \texttt{ATRAN} \citep{lord1992} for typical conditions at Gemini North and excluding any features for which the telluric spectrum exhibited transmission less than 0.90. The fluxes in these pixels were then compared to those expected for an absorption line whose equivalent width was 5 times larger than its uncertainty (see the Appendix~\ref{sec:app} for details). The pixels that survived this cut were visually inspected (to eliminate bad or hot pixels) and grouped by wavelength. Finally, central wavelengths and identifications were determined. We also examined the spectra from the observations on 04 Nov to identify detector artifacts and confirm the presence of intrinsic absorption lines. As mentioned above, the S/N ratios for the spectra from 04 Nov were substantially lower than those of the 06 Nov spectra and therefore were not included in our analysis. 

\begin{deluxetable}{ccccccccc}
\tabletypesize{\tiny}
\tabletypesize{\footnotesize}
\tablewidth{0pc}
\tablecaption{Literature compilation of atmospheric parameters for \protect\bdp. \label{tab:atmpar}}
\tablehead{
\colhead{\teff}&
\colhead{}&
\colhead{\logg}&
\colhead{}&
\colhead{\metal}&
\colhead{}&
\colhead{\vturb}&
\colhead{}&
\colhead{Ref.}\\
\colhead{K}&
\colhead{}&
\colhead{[cgs]}&
\colhead{}&
\colhead{}&
\colhead{}&
\colhead{\kmsec}&
\colhead{}&
\colhead{}}
\startdata
5510 && 2.60 && $-$2.71 && \nodata && 1 \\
5510 && 3.70 && $-$3.73 &&    1.30 && 2 \\
5510 && 3.70 && $-$3.68 &&    1.30 && 3 \\
5430 && 3.40 && $-$3.83 &&    1.30 && 4 \\
5430 && 3.40 && $-$3.88 &&    1.30 && 5 \\
5040 && 2.10 && $-$4.28 &&    1.35 && 6 \\
5430 && 3.40 && $-$4.07 &&    1.30 && 7 \\
5461 && 3.00 && $-$3.78 &&    1.90 && 8 \\
5351 && 3.12 && $-$3.96 &&    1.45 && 9\vv
\enddata
\tablenotetext{a}{Adopted in this work.}
\tablerefs{
1: \citet{carney2003};
2: \citet{ito2009};
3: \citet{takeda2011};
4: \citet{ito2013};
5: \citet{placco2014b};
6: \citet{roederer2014};
7: \citet{roederer2016};
8: \citet{jeong2023};
9: \citet{placco2024}.}
\end{deluxetable}

Identifications were made using line lists generated by the \texttt{linemake} code\footnote{\href{https://github.com/vmplacco/linemake}{https://github.com/vmplacco/linemake}} \citep{placco2021} and the Atomic Line List website\footnote{\href{https://linelist.pa.uky.edu/newpage/}{https://linelist.pa.uky.edu/newpage/}} \citep{vanhoof2018}. Equivalent widths (EWs) were determined for each absorption feature by fitting a lower-order polynomial to the fluxes within wavelength regions near the lines to estimate the continuum level, dividing the spectrum by the fit (to remove any residual variations in the previously normalized continuum), and directly integrating the fluxes within the line. The wavelengths, EWs, and line IDs are presented in Table \ref{tab:syn}. Both the $H$ and $K$-band spectra were shifted to the rest frame using the latest literature determination for the radial velocity of \bdp\, \citep[$-150.445 \pm 0.016$~\kmsec;][]{placco2024}. The velocity shift derived by measuring the wavelengths of the centroids of the H absorption lines in the spectra is $\sim 150.4$ km s$^{-1}$, in excellent agreement with this value and indicating that the wavelength calibration for the IGRINS-2 spectra is very accurate.

\begin{figure*}[ht!]
\includegraphics[width=0.5\textwidth]{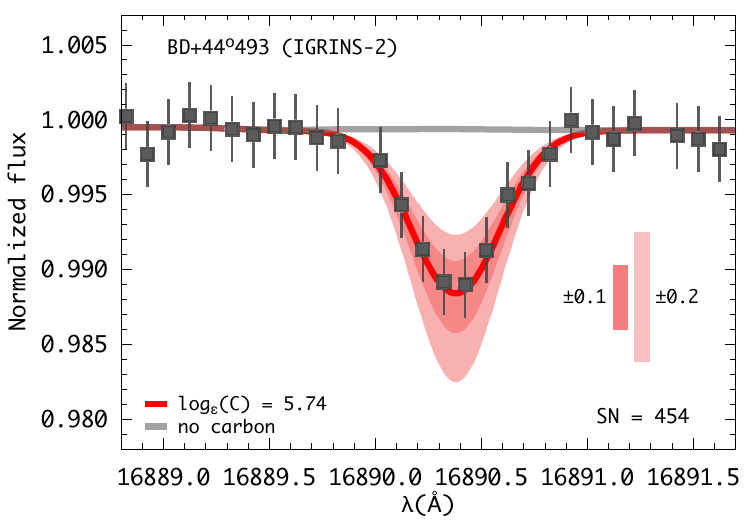}
\includegraphics[width=0.5\textwidth]{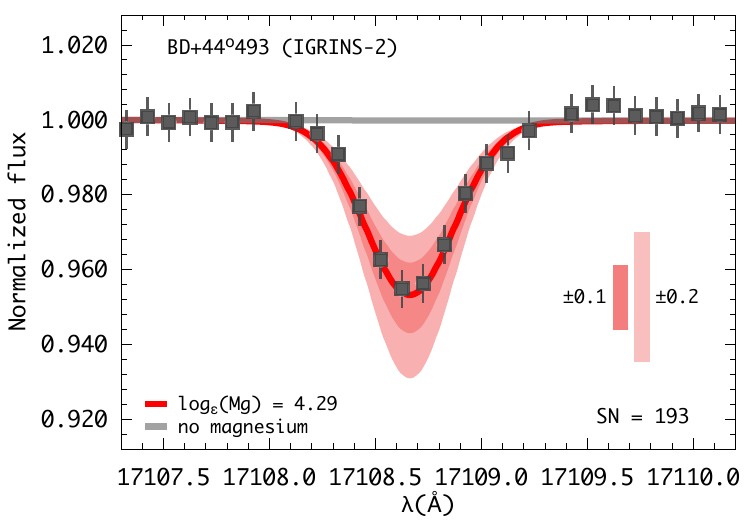}
\includegraphics[width=0.5\textwidth]{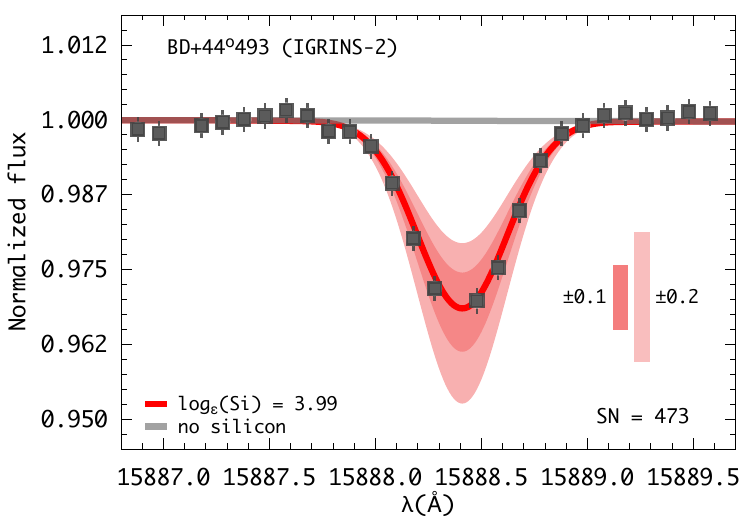}
\includegraphics[width=0.5\textwidth]{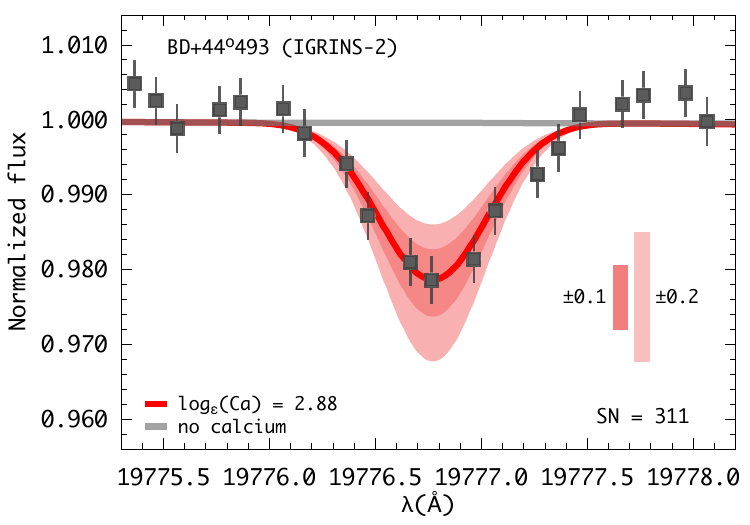}
\caption{Spectral synthesis for carbon (top left), magnesium (top right), silicon (bottom left), and calcium (bottom right). Filled squares connected by the black lines represent the IGRINS-2 spectrum, the red lines are the best fit, and the shaded regions represent $\pm0.1$ and $\pm0.2$~dex from the best-fit abundance. Shown as gray lines are the synthetic spectra without each element.}
\label{fig:synthesis}
\end{figure*}

\section{Abundance Determinations}
\label{sec:abund}

For the chemical abundance determinations, we used the stellar atmospheric parameters determined by \citet{placco2024}: \teff=5351~K, \logg=3.12, \metal=$-3.96$, and microturbulent velocity $\xi=1.45$~\kmsec. Over the years, several studies have determined these parameters using a combination of photometric and spectroscopic methods, and Table~\ref{tab:atmpar} lists the parameters found in the literature for \bdp. We chose to use the latest values provided by \citet{placco2024} not only because of the extremely high resolution and S/N of their data, but also due to the fact that recent studies of low-metallicity stars have been adopting the same methodology\footnote{The effective temperature is determined from color--\teff--\metal\, relations, the surface gravity (\logg) is determined using the measured parallax and fundamental relations, the \metal\, is determined spectroscopically from the EWs of \ion{Fe}{1} lines, and the microturbulent velocity is determined by minimizing the trend between the \ion{Fe}{1} abundances and their reduced EW.} in an attempt to homogenize samples across different studies and minimize systematic uncertainties \citep[e.g.][]{hansen2018,roederer2018,placco2020,placco2021,placco2023b,placco2025,placco2026}.

Abundances were determined through equivalent width analysis and spectral synthesis using the 2017 version of the \texttt{MOOG}\footnote{\href{https://github.com/alexji/moog17scat}{https://github.com/alexji/moog17scat}} code \citep{sneden1973,sobeck2011}, with one-dimensional plane-parallel model atmospheres without overshooting \citep{castelli2004} and assuming local thermodynamic equilibrium (LTE). The calculated abundances for the 10 absorption features identified in the spectrum of \bdp\ and their associated atomic data are given in Table~\ref{tab:syn}. Table~\ref{tab:syn} also gives the non-LTE (NLTE) corrections for the \ion{Mg}{1} and \ion{Si}{1} lines, calculated using the MPIA NLTE online database\footnote{\href{https://nlte.mpia.de/}{https://nlte.mpia.de/}}\citep{mpia}. Below, we provide details for each absorption feature measured in the IGRINS-2 data. 

\subsection{Carbon}

The carbon abundance was measured from the atomic \ion{C}{1} feature at 16890 \AA\, with a best-fit abundance of \eps{C}=$5.74 \pm 0.15$. The EW and synthesis values agree within $1\sigma$. The upper-left panel of Figure~\ref{fig:synthesis} shows the synthetic spectrum of the carbon line. The filled squares and black solid line represent the IGRINS-2 spectrum, the red line is the best fit, and the shaded regions represent $\pm0.1$ and $\pm0.2$~dex from the best-fit abundance, which is used to assess the abundance uncertainty. For reference, a synthetic spectrum without carbon (gray line) is shown. Despite the weak line strength (EW=4.23~m\AA), the line is nonetheless clearly detected due to the high quality of the IGRINS-2 data, with a statistical S/N $\sim 454$ in the region surrounding the \ion{C}{1} feature.

\subsection{Magnesium}

A total of seven \ion{Mg}{1} absorption features were measured in the H-band spectrum of \bdp, ranging from 14877 \AA\, to 17108 \AA. The calculated abundances derived from the individual features range from 4.29 to 4.56 in the synthesis case, with a median value of \eps{Mg}=$4.34 \pm 0.09$. The values calculated from the EW analysis agree within 0.17~dex for all but one line ($\lambda=14877$\AA) with $\Delta$\eps{Mg}=0.49. This difference is due to the fact that there are six \ion{Mg}{1} transitions between 14877.5 \AA\, and 14877.78 \AA\, \citep{pehlivan2017}, which can only be properly accounted for via spectral synthesis. The top right panel of Figure~\ref{fig:synthesis} shows the spectral synthesis of the \ion{Mg}{1} line at 17108\AA. The NLTE corrections range from 0.09 to 0.31 for six of the seven lines.

\subsection{Silicon}

One \ion{Si}{1} line was identified in the H-band at 15888 \AA\, with a calculated abundance of \eps{Si}=$3.99 \pm 0.10$ and a negligible NLTE correction ($\Delta$NLTE=$-0.02$). The spectral synthesis for this line is shown in the bottom left panel of Figure~\ref{fig:synthesis}, with S/N $\sim 473$. The spectral synthesis abundance agrees very well with the EW value.

\subsection{Calcium}

For the calcium abundance, one line in the K-band spectrum of \bdp\, at 19776 \AA\, was measured, with a calculated abundance of \eps{Ca}=$2.88 \pm 0.15$. The lower-right panel of Figure~\ref{fig:synthesis} shows the spectral synthesis of this line. 
The EW abundance is 0.10~dex higher than the spectral synthesis value, which is within the expected uncertainties.

\begin{deluxetable}{lcccCcc}
\tabletypesize{\tiny}
\tabletypesize{\footnotesize}
\tablewidth{0pc}
\tablecaption{\label{tab:abun} 1D LTE Abundances from the literature and recalculated in this work.}
\tablehead{
\colhead{Species}&
\colhead{$\log\epsilon$\,(X)}&
\colhead{$\log\epsilon$\,(X)\vv}&
\colhead{$\sigma$}&
\colhead{$N$}&
\colhead{$\lambda$}&
\colhead{Ref.}\\
\colhead{}&
\colhead{Literature}&
\colhead{This work}&
\colhead{}&
\colhead{}&
\colhead{}&
\colhead{}
}
\startdata
\ion{C}{1}  & 5.74 & 5.74    & 0.15 &   1 & NIR & 1 \\
            & 5.78 & 5.77    & 0.20 &   2 & NUV & 5 \\
            & 6.07 & 5.91    & 0.20 &   1 & NIR & 6 \\
        CH  & 5.87 & 5.87    & 0.10 &   1 & Opt & 2 \\
            & 5.80 & 5.83    & 0.20 & 105 & Opt & 4 \\
            & 5.95 & 5.82    & 0.37 &   1 & Opt & 7 \\
\hline      
\ion{Mg}{1} & 4.34 & 4.34    & 0.09 &   7 & NIR & 1 \\
            & 4.25 & 4.25    & 0.12 &   8 & Opt & 2 \\ 
            & 4.23 & 4.54    & 0.50 &   1 & NUV & 3 \\
            & 4.23 & 4.17    & 0.21 &   8 & Opt & 7 \\
            & 4.51 & 4.47    & 0.10 &   3 & NIR & 8 \\
\hline      
\ion{Si}{1} & 3.99 & 3.99    & 0.10 &   1 & NIR & 1 \\
            & 3.90 & 3.90    & 0.10 &   1 & Opt & 2 \\
            & 3.78 & 4.07    & 0.24 &  10 & NUV & 3 \\
            & 4.17 & 4.00    & 0.26 &   1 & Opt & 7 \\
            & 4.18 & 4.11    & 0.11 &   5 & NIR & 9 \\
\hline      
\ion{Ca}{1} & 2.88 & 2.88    & 0.15 &   1 & NIR & 1 \\
            & 2.73 & 2.73    & 0.11 &   7 & Opt & 2 \\
            & 2.82 & 2.77    & 0.11 &   4 & Opt & 7 \\
\enddata
\tablenotetext{a}{Abundances calculated using the stellar parameters from \citet{placco2024}: \teff=5351~K, \logg=3.12, \metal=$-3.96$, and $\xi=1.45$~\kmsec.}
\tablerefs{\\
1: This work;
2: \citet{placco2024};
3: \citet{roederer2016};
4: \citet{aoki2015};
5: \citet{placco2014b};
6: \citet{takeda2013};
7: \citet{ito2013};
8: \citet{aoki2025};
9: \citet{aoki2022}.
}
\end{deluxetable}

\subsection{Redetermining Chemical Abundances from the Literature}
\label{sec:redeter}

We redetermined the chemical abundances of carbon, magnesium, silicon, and calcium for \bdp\ given in the literature, using the atmospheric parameters determined by \citet{placco2024}. This helps homogenize the abundances and produce a more consistent basis for comparison (see Section~\ref{sec:disc}). The results of this exercise are summarized in Table~\ref{tab:abun}, and we provide a few details for each element below.

\paragraph{Carbon} From \citet{takeda2013}, we used the EW for the \ion{C}{1} line at 10691\AA\ line, to derive a carbon abundance of \eps{C}=5.91. In the case of \citet{aoki2015}, all 105 EW values from the CH $B-X$ and $A-X$ absorption features were used, giving an average abundance of \eps{C}=$5.83\pm0.06$. Carbon abundances were also derived from synthesis of the spectra published in \citet{ito2013} and \citet{placco2014b}, using the CH G-band around 4303\AA\, (which yielded \eps{C}=5.82) and the 2478\AA/2967\AA\, \ion{C}{1} lines (which gave \eps{C}=5.77), respectively.

\paragraph{Magnesium} Abundances were calculated from the EW values given by \citet{ito2013} (8 lines - \eps{Mg}=$4.17\pm0.12$), \citet{roederer2016} (2025\AA\, - \eps{Mg}=4.54), and \citet{aoki2025} (3 lines - \eps{Mg}=$4.47\pm0.12$).

\paragraph{Silicon} Abundances were calculated from EWs given by \citet{ito2013} (3905\AA\, - \eps{Mg}=4.00), \citet{roederer2016} (10 lines - \eps{Mg}=$4.07\pm0.24$), and \citet{aoki2025} (5 lines - \eps{Mg}=$4.11\pm0.14$).

\paragraph{Calcium} Although a calcium abundance is published in \citet{ito2013}, we were not able to find the line-by-line EW values. From the Subaru spectrum, we measured the EWs of four lines (3644\AA, 4318\AA, 4434\AA, and 4454\AA), yielding an average abundance \eps{Ca}=$2.77\pm0.05$.

\subsection{Systematic Uncertainties}

Systematic uncertainties arising from variations in stellar atmospheric parameters were assessed using abundances determined via EW analysis. Table~\ref{tab:sys} shows the variations in chemical abundances for the four elements identified in the IGRINS-2 data when the atmospheric parameters are changed by $+$150~K for \teff, $+$0.30~dex for \logg, and $+$0.30~km\,s$^{-1}$ for $\xi$, to roughly account for the variation in parameters seen in Table~\ref{tab:atmpar}. The $\sigma$ values are the standard errors of the abundances determined in this work. For each element, the total uncertainty ($\sigma_{\rm tot}$) is calculated from the quadratic sum of the individual error estimates.

\begin{deluxetable}{lrcccc}[!ht]
\tabletypesize{\small}
\tabletypesize{\footnotesize}
\tablewidth{0pc}
\tablecaption{Systematic Abundance Uncertainties for \protect \bdp. \label{tab:sys}}
\tablehead{
\colhead{Ion}&
\colhead{$\Delta$\teff}&
\colhead{$\Delta$\logg}&
\colhead{$\Delta\xi$}&
\colhead{$\sigma$}&
\colhead{$\sigma_{\rm tot}$\tablenotemark{a}}\\
\colhead{}&
\colhead{$+$150\,K}&
\colhead{$+$0.30 dex}&
\colhead{$+$0.30 \kmsec}&
\colhead{}&
\colhead{}}
\startdata
\ion{C}{1}  & $-$0.09 &  0.12 &    0.00 &  0.15 &  0.21 \\  
\ion{Mg}{1} &    0.06 &  0.01 &    0.01 &  0.09 &  0.11 \\  
\ion{Si}{1} &    0.08 &  0.03 &    0.00 &  0.10 &  0.13 \\  
\ion{Ca}{1} &    0.13 &  0.01 &    0.00 &  0.15 &  0.20 \\  
\enddata
\tablenotetext{a}{Calculated from the quadratic sum of the individual error estimates.}
\end{deluxetable}

\begin{figure*}[ht!]
\includegraphics[width=1.0\textwidth]{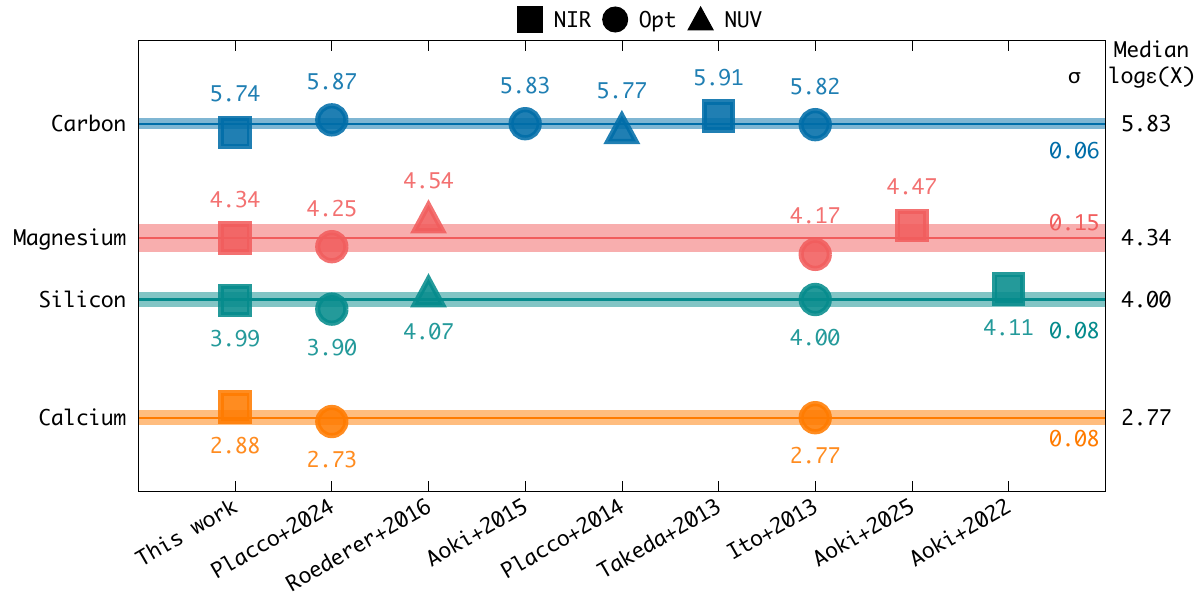}
\caption{Abundances for \protect\bdp\, from the literature, re-calculated using the model atmosphere from \citet{placco2024}. Solid lines represent the median abundances, and shaded areas represent 1$\sigma$. Different symbols refer to the wavelength regimes highlighted at the top.}
\label{fig:ab_comp}
\end{figure*}

\section{Discussion}
\label{sec:disc}

The chemical abundances for \bdp\ calculated from the IGRINS-2 data add valuable information to its already extensive abundance inventory. A compilation of previous abundance determinations for C, Mg, Si, and Ca for \bdp\, is given in Table \ref{tab:abun}, which includes the re-calculated abundances for all the literature references (see Section~\ref{sec:redeter} for details) using the model atmosphere from \citet{placco2024}. Figure~\ref{fig:ab_comp} shows a comparison of the recalculated abundances from the nine studies listed in Table~\ref{tab:abun}, along with the median value (solid horizontal lines) and standard deviation (colored shaded areas). The values calculated in this work (from single lines for \ion{C}{1}, \ion{Si}{1}, and \ion{Ca}{1}, and the average of seven lines for \ion{Mg}{1}) agree remarkably well with the re-calculated values from the literature, with $\sigma\leq0.08$~dex for carbon, silicon, and calcium. In the specific case of magnesium, the abundances range from \eps{Mg}=4.17 to 4.54. It is worth noting that \citet{roederer2016} estimates $\sigma=0.50$ for the single \ion{Mg}{1} line at 2025\AA\, in the COS spectrum. For the recalculated \ion{Mg}{1} abundances for both \citet{ito2013} and \citet{aoki2025}, the standard deviation values are $\sigma=0.12$.

To assess whether these updated abundances could change the conclusions reached by \citet{placco2024} about the progenitor population of \bdp, we repeated their analysis using the \texttt{starfit}\footnote{\href{https://starfit.org/}{https://starfit.org/}} code, which contains nucleosynthesis yields for 16,800 metal-free Population~III faint-supernova \citep{heger2002,heger2010}, updating the abundances of C, Mg, Si, and Ca. The code finds the best-fit model (in terms of progenitor mass and explosion energy) for each of the 10,000 re-sampled chemical abundance patterns for \bdp\, \citep[see][for further details on the methodology]{placco2016b}. There were no changes in the most frequent progenitor models, which are in the 20.5-21.5\,M$_\odot$ mass range, with low explosion energies within $0.3-1.2 \times 10^{51}$\,erg. In addition, using the new average \xfe{C}=$+1.36$ and \xfe{Mg}=$+0.70$, we calculate a \abund{Mg}{C}=$-0.66$, which is still consistent with the ``mono-enriched'' classification based on the criteria outlined in \citet{hartwig2018}. More recently, \citet{hartwig2023} developed a data-driven method to classify UMP stars as mono- or multi-enriched using support vector machines. We used the {\texttt{emu-c}} code\footnote{\href{https://gitlab.com/thartwig/emu-c}{https://gitlab.com/thartwig/emu-c}}, which takes as input abundances of C, O, Na, Mg, Al, Si, Ca, Cr, Mn, Fe, Co, Ni, Zn, and Ba. Based on the LTE abundances in Table~\ref{tab:abun} and \citet{placco2024}, the probability that \bdp\, is mono-enriched is $61\pm11$\%, compared with $59\pm12$\% when using only the abundances from \citet{placco2024}.

\section{Conclusions}
\label{sec:conc}

In this work, we analyzed the IGRINS-2 NIR spectra of the CEMP star \bdp. The extremely high S/N allowed for the identification of 10 weak absorption lines, including the first measurement of a \ion{C}{1} feature in the $H$-band for a UMP star. Comparison with the values derived in this study indicates that our values agree well with those reported in the literature and have uncertainties that are similar to, or in some cases smaller than, those reported. The updated abundances confirm previous claims that \bdp\, is a bona fide second-generation star, potentially formed from a gas cloud enriched by a single low-energy $\sim20$\,M$_\odot$ supernova in the early universe.

These observations demonstrate the usefulness of the $H$ and $K$ bands for abundance determinations. Lines as weak as a few m\AA\, can be reliably detected at a resolution of $\sim 45,000$ with IGRINS-2 on Gemini North in a few minutes of on-source exposure time for objects with $K\sim 7$. 
At $K\sim 12$, the IGRINS-2 exposure time calculator predicts a S/N$\sim ~100$ in 1 hour for the $K$ band. 
The combination of the large simultaneous NIR wavelength coverage and high resolution provided by IGRINS-2, the 8-meter aperture of Gemini, and the favorable NIR atmospheric transmission at Maunakea, along with the simplicity of the observations and observation preparations and the straightforward nature of the data reduction, should make IGRINS-2 on Gemini North a workhorse instrument for stellar abundance determinations.

\begin{acknowledgments}

The authors would like to thank the anonymous referee for their comments and suggestions; Anish Amarsi for a detailed review of the text and discussion of future work on 3D/NLTE abundances; Thomas Nordlander for suggestions on an earlier version of the manuscript.
The work of V.M.P. and W.D.V. is supported by NOIRLab, which is managed by the Association of Universities for Research in Astronomy (AURA) under a cooperative agreement with the U.S. National Science Foundation. 
Based on observations obtained at the International Gemini Observatory (Program ID: GN-2025B-Q-415), a program of NSF NOIRLab, which is managed by the Association of Universities for Research in Astronomy (AURA) under a cooperative agreement with the U.S. National Science Foundation on behalf of the Gemini Observatory partnership: the U.S. National Science Foundation (United States), National Research Council (Canada), Agencia Nacional de Investigaci\'{o}n y Desarrollo (Chile), Ministerio de Ciencia, Tecnolog\'{i}a e Innovaci\'{o}n (Argentina), Minist\'{e}rio da Ci\^{e}ncia, Tecnologia, Inova\c{c}\~{o}es e Comunica\c{c}\~{o}es (Brazil), and Korea Astronomy and Space Science Institute (Republic of Korea).
This research has made use of NASA's Astrophysics Data System Bibliographic Services; the arXiv pre-print server operated by Cornell University; and the {\texttt{SIMBAD}} database hosted by the Strasbourg Astronomical Data Center.

\end{acknowledgments}

\software{
{\texttt{Astropy}}\,\citep{astropy2013,astropy2018}, 
{\texttt{awk}}\,\citep{awk}, 
{\texttt{emu-c}}\,\citep{hartwig2023},
{\texttt{gnuplot}}\,\citep{gnuplot}, 
{\texttt{linemake}}\,\citep{placco2021,placco2021a},
{\texttt{MOOG}}\,\citep{sneden1973},  
{\texttt{NOIRLab IRAF}}\,\citep{tody1986,tody1993,fitzpatrick2025}, 
{\texttt{numpy}}\,\citep{numpy}, 
{\texttt{sed}}\,\citep{sed},
{\texttt{Spectroscopy Made Harder (SMHr)}}\,\citep{casey2014},
{\texttt{starfit}}\,\citep{heger2002,heger2010},
{\texttt{xtellcor}}\,\citep{vacca2003}.
}

\facilities{Gemini:Gillett (IGRINS-2)}

\newpage

\appendix

\section{Line Identification Criterion}
\label{sec:app}

In this appendix, we present details on the equations used to detect absorption lines in our spectra. The procedure is based on the results of \citet{cayrel1988} and \citet{Landman1982}. Because the former reference is not available electronically, we provide a derivation of those results here.

We assume that the observed flux in the vicinity of a feature in an IGRINS-2 spectrum can be modeled as a continuum with a Gaussian absorption line 
\[
I_\lambda = I_c - I_o\exp\!\left[-\dfrac{(\lambda - \lambda_0)^2}{2\Delta^2}\right]
\]
and therefore
\[
\dfrac{I_\lambda}{I_c} = 1 - \dfrac{I_o}{I_c}\exp\!\left[-\dfrac{(\lambda - \lambda_0)^2}{2\Delta^2}\right]
\]
where $I_c$ is the continuum, $I_0$ is the height of the Gaussian line, $\lambda_0$ is the central wavelength of the line, and $\Delta$ is the Gaussian standard deviation. The equivalent width of the absorption line, $W_\lambda$, is then given by
\[
W_\lambda = \int\frac{I_o}{I_c}\exp\!\left[-\frac{(\lambda - \lambda_0)^2}{2\Delta^2}\right]d\lambda
= \frac{\sqrt{\pi}}{2\sqrt{\ln 2}}\cdot\frac{I_o}{I_c}\cdot\text{FWHM}
= 1.604 \cdot \frac{I_o}{I_c}\cdot\text{FWHM}
\]
where $\Delta = \dfrac{\text{FWHM}}{2\sqrt{2\ln 2}}$ for a Gaussian line. 
\citet{Landman1982} considered the uncertainties when fitting the function
\[
y = A\exp\!\left[-\frac{(x-B)^2}{C^2}\right]
\]
to continuum-subtracted spectral data. They found that the uncertainty in the fitted value of $A$ is given by

\[
\sigma_A = \left[\frac{9}{2\pi}\right]^{1/4}\left[\frac{\delta x}{C}\right]^{1/2}\varepsilon 
\]
where $\delta x$ is the pixel size and $\varepsilon$ is the noise in the pixel values, assumed to be the same for all pixels in the vicinity of the line. In our formulation, $A = \frac{I_o}{I_c}$, $C = \sqrt{2}\,\Delta = \frac{\text{FWHM}}{2\sqrt{\ln 2}}$, $\delta x = \delta \lambda$, and we have

\[
\sigma_{I_o/I_c} = \left[\frac{9}{4\pi}\right]^{1/4}\cdot \left[\frac{\delta \lambda}{\Delta}\right]^{1/2}\cdot\varepsilon
= 0.920\cdot \left(\frac{\delta \lambda}{\Delta}\right)^{1/2}\varepsilon ~~.
\]
This expression is the equivalent of equation 3 in \citet{cayrel1988}. In terms of the FWHM, we have

\[
\sigma_{I_o/I_c} = \left[\frac{18\ln 2}{\pi}\right]^{1/4}\cdot\left[\frac{\delta \lambda}{\text{FWHM}}\right]^{1/2}\cdot\varepsilon = 1.412 \cdot \left[\frac{\delta \lambda}{\text{FWHM}}\right]^{1/2}\cdot\varepsilon~~.
\]
From the definition of $W_\lambda$, and with the assumption that $\sigma_{\text{FWHM}} = 0$ (which is strictly true only for an unresolved line),

\[
\sigma_{W_\lambda} = \frac{\sqrt{\pi}}{2\sqrt{\ln 2}}\cdot\sigma_{I_o/I_c}\cdot\text{FWHM}
= \left[\frac{9\pi}{8\ln 2}\right]^{1/4}\cdot\left[\delta \lambda\cdot\text{FWHM}\right]^{1/2}\cdot\varepsilon
= 1.503\cdot\left[\delta \lambda\cdot\text{FWHM}\right]^{1/2}\cdot\varepsilon
\]

which reproduces equation 6' in \citet{cayrel1988}. For an unresolved line, $\text{FWHM} = \frac{\lambda_0}{R}$ and then 
\[
\sigma_{W_\lambda} 
= 1.503\cdot\left[\frac{\delta \lambda\cdot\lambda_0}{R}\right]^{1/2}\cdot\varepsilon ~~~.
\]

The noise in the continuum-normalized pixel values, $\varepsilon$, is given by the uncertainty in the value of $\frac{I_\lambda}{I_c}$ in the vicinity of the line
\[
\varepsilon = \sigma_{I_\lambda/I_c}
\approx \frac{\sigma_{I_\lambda}}{I_c} \approx \left[S/N\right]^{-1}
\]
where $S/N$ is the signal-to-noise ratio in the spectrum in the vicinity of the line, and we have made the (somewhat unrealistic) assumption of no uncertainty in the determination of $I_c$.
Then we\footnote{We note that \citet{casey2014} gives a similar expression for the minimum detectable EW, $W^{min}_{\lambda}$ (for which, presumably, $W^{min}_{\lambda} = \sigma_{W_\lambda}$), based on a ``revised version" of the equation of \citet{cayrel1988}
\[
W^{min}_{\lambda} = \left[1.5 \cdot \text{FWHM} \cdot \delta\lambda\right]^{1/2} \cdot \left[S/N\right]^{-1} ~~,
\]
but provides no details on the derivation.}
have

\[
\sigma_{W_\lambda} 
= 1.503\cdot\left[\delta \lambda\cdot\text{FWHM}\right]^{1/2}\cdot\left[S/N\right]^{-1}
= 1.503\cdot\left[\frac{\delta \lambda\cdot\lambda_0}{R}\right]^{1/2}\cdot\left[S/N\right]^{-1} ~~.
\]

For example, we can use this equation to estimate the uncertainty in the measured equivalent width of the C I line. For the parameters of the IGRINS-2 observations in the vicinity of the line, we find $\sigma_{W_\lambda} \sim 0.6\ \text{m\AA}$, which is close to the empirical value derived from numerically integrating over the spectrum. 

Following \citet{casey2014}, one can now use the estimate derived from this equation as a threshold for determining whether a putative line (found either automatically or by eye) is real. If we assert that the equivalent width of a feature must be larger than $f\cdot\sigma_{W_{\lambda}}$ for a believable detection and identification, then we can derive a limit on the normalized flux at the center of the putative unresolved line:

\[
W_{\lambda}= 1.064\cdot\left(\frac{\lambda_0}{R}\right)\cdot\left(\frac{I_o}{I_c}\right) >  f\cdot 1.503\cdot\left[\frac{\delta \lambda\cdot\lambda}{R}\right]^{1/2}\cdot\left[S/N\right]^{-1}
\]
so that 
\[
\frac{I_o}{I_c} > 
\frac{1.413\cdot f}{(S/N)}\cdot \left[\frac{\delta\lambda}{\text{FWHM}}\right]^{1/2}
= \frac{1.412 \cdot f}{(S/N)} \left[\frac{\delta\lambda\cdot R}{\lambda_0}\right]^{1/2}
\]
At the center of the absorption line, $I_{\lambda_0} = I_c - I_0$ and therefore 
\[
\frac{I_0}{I_c} = \left(1 - \frac{I_{\lambda_0}}{I_c}\right) > 
\frac{1.412 \cdot f}{(S/N)} \left[\frac{\delta\lambda\cdot R}{\lambda_0}\right]^{1/2}
\]
The limit on the observed flux at the center of the putative absorption line, in terms of purely observed and instrumental quantities, is then given by 
\[
\frac{I_{\lambda_0}}{I_c} < 1 - 
\frac{1.412 \cdot f}{(S/N)} \left[\frac{\delta\lambda\cdot R}{\lambda_0}\right]^{1/2}
\]
An additional criterion for the feature width could also be implemented to distinguish between a bad pixel and a real line.

Finally, we note that the definition of $W_\lambda$ incorporates an integration over all wavelengths. In practice, the integral is computed over a wavelength range from $\lambda_0 - L$ to $\lambda_0 + L$. In this case, it is straightforward to show that the measured equivalent width is given by
\[
W_\lambda^{meas} = 
\frac{\sqrt{\pi}}{2\sqrt{\ln 2}} \cdot \frac{I_o}{I_c} \cdot \text{FWHM} \cdot \text{erf}\left(\frac{2L\sqrt{\ln 2}}{\text{FWHM}}\right)
= 1.064\cdot\frac{I_o}{I_c} \cdot \text{FWHM} \cdot \text{erf}\!\left(\frac{1.665\,L}{\text{FWHM}}\right)
\]
Therefore, the relation between the measured value and the true value is
\[
W_{\lambda}^{true} = \dfrac{W_\lambda^{meas}}{\gamma}
\]
where 
\[ 
\gamma = \text{erf}\left(\frac{2L\sqrt{\text{ln~2}}}{\text{FWHM}}\right) ~~.
\]
If $L \approx \text{FWHM}/2$ then $\gamma = 0.761$ and the measured equivalent width must be increased by a factor of 1.314 to obtain the true value. If $L \sim \text{FWHM}$, as it is for our measurements, then $\gamma \approx 0.98$.

\newpage

\bibliographystyle{aasjournal}
\bibliography{bibliografia}

@ARTICLE{aoki2007,
   author = {{Aoki}, W. and {Beers}, T.~C. and {Christlieb}, N. and {Norris}, J.~E. and 
	{Ryan}, S.~G. and {Tsangarides}, S.},
    title = "{Carbon-enhanced Metal-poor Stars. I. Chemical Compositions of 26 Stars}",
  journal = {\apj},
   eprint = {arXiv:astro-ph/0609702},
     year = 2007,
    month = jan,
   volume = 655,
    pages = {492-521},
      doi = {10.1086/509817},
   adsurl = {http://adsabs.harvard.edu/abs/2007ApJ...655..492A}
}

@ARTICLE{york2000,
   author = {{York}, D.~G. and {Adelman}, J. and {Anderson}, Jr., J.~E. and 
	{Anderson}, S.~F. and {Annis}, J. and {Bahcall}, N.~A. and {Bakken}, J.~A. and 
	{Barkhouser}, R. and {Bastian}, S. and {Berman}, E. and {Boroski}, W.~N. and 
	{Bracker}, S. and {Briegel}, C. et al.},
    title = "{The Sloan Digital Sky Survey: Technical Summary}",
  journal = {\aj},
   eprint = {arXiv:astro-ph/0006396},
     year = 2000,
    month = sep,
   volume = 120,
    pages = {1579-1587},
      doi = {10.1086/301513},
   adsurl = {http://adsabs.harvard.edu/abs/2000AJ....120.1579Y}
}

@ARTICLE{christlieb2008,
   author = {{Christlieb}, N. and {Sch{\"o}rck}, T. and {Frebel}, A. and 
	{Beers}, T.~C. and {Wisotzki}, L. and {Reimers}, D.},
    title = "{The stellar content of the Hamburg/ESO survey. IV. Selection of candidate metal-poor stars}",
  journal = {\aap},
   eprint = {arXiv:0804.1520},
     year = 2008,
    month = jun,
   volume = 484,
    pages = {721-732},
      doi = {10.1051/0004-6361:20078748},
   adsurl = {http://adsabs.harvard.edu/abs/2008A%26A...484..721C}
}

@PHDTHESIS{sneden1973,
   author = {{Sneden}, C.~A.},
    title = "{Carbon and Nitrogen Abundances in Metal-Poor Stars.}",
   school = {The University of Texas at Austin.},
     year = 1973,
   adsurl = {http://adsabs.harvard.edu/abs/1973PhDT.......180S}
}

@ARTICLE{castelli2004,
   author = {{Castelli}, F. and {Kurucz}, R.~L.},
    title = "{New Grids of ATLAS9 Model Atmospheres}",
  journal = {ArXiv Astrophysics e-prints},
   eprint = {arXiv:astro-ph/0405087},
     year = 2004,
    month = may,
   adsurl = {http://adsabs.harvard.edu/abs/2004astro.ph..5087C}
}

@ARTICLE{bromm2004,
   author = {{Bromm}, V. and {Larson}, R.~B.},
    title = "{The First Stars}",
  journal = {\araa},
   eprint = {arXiv:astro-ph/0311019},
     year = 2004,
    month = sep,
   volume = 42,
    pages = {79-118},
      doi = {10.1146/annurev.astro.42.053102.134034},
   adsurl = {http://adsabs.harvard.edu/abs/2004ARA%26A..42...79B}
}

@ARTICLE{ito2009,
   author = {{Ito}, H. and {Aoki}, W. and {Honda}, S. and {Beers}, T.~C.},
    title = "{BD$+$44$^{o}$493: A Ninth Magnitude Messenger from the Early Universe; Carbon Enhanced and Beryllium Poor}",
  journal = {\apjl},
archivePrefix = "arXiv",
   eprint = {0905.0950},
     year = 2009,
    month = jun,
   volume = 698,
    pages = {L37-L41},
      doi = {10.1088/0004-637X/698/1/L37},
   adsurl = {http://adsabs.harvard.edu/abs/2009ApJ...698L..37I}
}

@ARTICLE{cayrel2004,
   author = {{Cayrel}, R. and {Depagne}, E. and {Spite}, M. and {Hill}, V. and 
	{Spite}, F. and {Fran{\c c}ois}, P. and {Plez}, B. and {Beers}, T. and 
	{Primas}, F. and {Andersen}, J. and {Barbuy}, B. and {Bonifacio}, P. and 
	{Molaro}, P. and {Nordstr{\"o}m}, B.},
    title = "{First stars V - Abundance patterns from C to Zn and supernova yields in the early Galaxy}",
  journal = {\aap},
   eprint = {arXiv:astro-ph/0311082},
     year = 2004,
    month = mar,
   volume = 416,
    pages = {1117-1138},
      doi = {10.1051/0004-6361:20034074},
   adsurl = {http://adsabs.harvard.edu/abs/2004A%26A...416.1117C}
}

@ARTICLE{sobeck2011,
   author = {{Sobeck}, J.~S. and {Kraft}, R.~P. and {Sneden}, C. and {Preston}, G.~W. and 
	{Cowan}, J.~J. and {Smith}, G.~H. and {Thompson}, I.~B. and 
	{Shectman}, S.~A. and {Burley}, G.~S.},
    title = "{The Abundances of Neutron-capture Species in the Very Metal-poor Globular Cluster M15: A Uniform Analysis of Red Giant Branch and Red Horizontal Branch Stars}",
  journal = {\aj},
archivePrefix = "arXiv",
   eprint = {1103.1008},
 primaryClass = "astro-ph.SR",
     year = 2011,
    month = jun,
   volume = 141,
      eid = {175},
    pages = {175},
      doi = {10.1088/0004-6256/141/6/175},
   adsurl = {http://adsabs.harvard.edu/abs/2011AJ....141..175S}
}

@ARTICLE{cowan2002,
   author = {{Cowan}, J.~J. and {Sneden}, C. and {Burles}, S. and {Ivans}, I.~I. and 
	{Beers}, T.~C. and {Truran}, J.~W. and {Lawler}, J.~E. and {Primas}, F. and 
	{Fuller}, G.~M. and {Pfeiffer}, B. and {Kratz}, K.-L.},
    title = "{The Chemical Composition and Age of the Metal-poor Halo Star BD +17{\deg}3248}",
  journal = {\apj},
   eprint = {arXiv:astro-ph/0202429},
     year = 2002,
    month = jun,
   volume = 572,
    pages = {861-879},
      doi = {10.1086/340347},
   adsurl = {http://adsabs.harvard.edu/abs/2002ApJ...572..861C}
}

@ARTICLE{norris2013,
   author = {{Norris}, J.~E. and {Bessell}, M.~S. and {Yong}, D. and {Christlieb}, N. and 
	{Barklem}, P.~S. and {Asplund}, M. and {Murphy}, S.~J. and {Beers}, T.~C. and 
	{Frebel}, A. and {Ryan}, S.~G.},
    title = "{The Most Metal-poor Stars. I. Discovery, Data, and Atmospheric Parameters}",
  journal = {\apj},
archivePrefix = "arXiv",
   eprint = {1208.2999},
 primaryClass = "astro-ph.GA",
     year = 2013,
    month = jan,
   volume = 762,
      eid = {25},
    pages = {25},
      doi = {10.1088/0004-637X/762/1/25},
   adsurl = {http://adsabs.harvard.edu/abs/2013ApJ...762...25N}
}

@ARTICLE{heger2010,
   author = {{Heger}, A. and {Woosley}, S.~E.},
    title = "{Nucleosynthesis and Evolution of Massive Metal-free Stars}",
  journal = {\apj},
archivePrefix = "arXiv",
   eprint = {0803.3161},
     year = 2010,
    month = nov,
   volume = 724,
    pages = {341-373},
      doi = {10.1088/0004-637X/724/1/341},
   adsurl = {http://adsabs.harvard.edu/abs/2010ApJ...724..341H}
}

@ARTICLE{ito2013,
   author = {{Ito}, H. and {Aoki}, W. and {Beers}, T.~C. and {Tominaga}, N. and 
	{Honda}, S. and {Carollo}, D.},
    title = "{Chemical Analysis of the Ninth Magnitude Carbon-enhanced
    Metal-poor Star BD+44:493}",
  journal = {\apj},
archivePrefix = "arXiv",
   eprint = {1306.3614},
 primaryClass = "astro-ph.SR",
     year = 2013,
    month = aug,
   volume = 773,
      eid = {33},
    pages = {33},
      doi = {10.1088/0004-637X/773/1/33},
   adsurl = {http://adsabs.harvard.edu/abs/2013ApJ...773...33I}
}

@ARTICLE{roederer2012b,
   author = {{Roederer}, I.~U. and {Lawler}, J.~E.},
    title = "{Detection of Elements at All Three r-process Peaks in the Metal-poor Star HD 160617}",
  journal = {\apj},
archivePrefix = "arXiv",
   eprint = {1204.3901},
 primaryClass = "astro-ph.SR",
     year = 2012,
    month = may,
   volume = 750,
      eid = {76},
    pages = {76},
      doi = {10.1088/0004-637X/750/1/76},
   adsurl = {http://adsabs.harvard.edu/abs/2012ApJ...750...76R}
}

@ARTICLE{sneden1998,
   author = {{Sneden}, C. and {Cowan}, J.~J. and {Burris}, D.~L. and {Truran}, J.~W.
	},
    title = "{Hubble Space Telescope Observations of Neutron-Capture Elements in Very Metal Poor Stars}",
  journal = {\apj},
     year = 1998,
    month = mar,
   volume = 496,
    pages = {235},
      doi = {10.1086/305366},
   adsurl = {http://adsabs.harvard.edu/abs/1998ApJ...496..235S}
}

@ARTICLE{takeda2013,
   author = {{Takeda}, Y. and {Takada-Hidai}, M.},
    title = "{Carbon Abundances of Metal-Poor Stars Determined from the C I 1.068-1.069{$\mu$}m Lines}",
  journal = {\pasj},
archivePrefix = "arXiv",
   eprint = {1302.3313},
 primaryClass = "astro-ph.SR",
     year = 2013,
    month = jun,
   volume = 65,
    pages = {65},
      doi = {10.1093/pasj/65.3.65},
   adsurl = {http://adsabs.harvard.edu/abs/2013PASJ...65...65T}
}

@ARTICLE{carney2003,
   author = {{Carney}, B.~W. and {Latham}, D.~W. and {Stefanik}, R.~P. and 
	{Laird}, J.~B. and {Morse}, J.~A.},
    title = "{Spectroscopic Binaries, Velocity Jitter, and Rotation in Field Metal-poor Red Giant and Red Horizontal-Branch Stars}",
  journal = {\aj},
     year = 2003,
    month = jan,
   volume = 125,
    pages = {293-321},
      doi = {10.1086/345386},
   adsurl = {http://adsabs.harvard.edu/abs/2003AJ....125..293C}
}

@ARTICLE{placco2014b,
   author = {{Placco}, V.~M. and {Beers}, T.~C. and {Roederer}, I.~U. and 
	{Cowan}, J.~J. and {Frebel}, A. and {Filler}, D. and {Ivans}, I.~I. and 
	{Lawler}, J.~E. and {Schatz}, H. and {Sneden}, C. and {Sobeck}, J.~S. and 
	{Aoki}, W. and {Smith}, V.~V.},
    title = "{Hubble Space Telescope Near-ultraviolet Spectroscopy of the Bright CEMP-no Star BD$+44^o493$}",
  journal = {\apj},
archivePrefix = "arXiv",
   eprint = {1406.0538},
 primaryClass = "astro-ph.SR",
     year = 2014,
    month = jul,
   volume = 790,
      eid = {34},
    pages = {34},
      doi = {10.1088/0004-637X/790/1/34},
   adsurl = {http://adsabs.harvard.edu/abs/2014ApJ...790...34P}
}

@ARTICLE{roederer2014,
   author = {{Roederer}, I.~U. and {Preston}, G.~W. and {Thompson}, I.~B. and 
	{Shectman}, S.~A. and {Sneden}, C. and {Burley}, G.~S. and {Kelson}, D.~D.
	},
    title = "{A Search for Stars of Very Low Metal Abundance. VI. Detailed Abundances of 313 Metal-poor Stars}",
  journal = {\aj},
archivePrefix = "arXiv",
   eprint = {1403.6853},
 primaryClass = "astro-ph.SR",
     year = 2014,
    month = jun,
   volume = 147,
      eid = {136},
    pages = {136},
      doi = {10.1088/0004-6256/147/6/136},
   adsurl = {http://adsabs.harvard.edu/abs/2014AJ....147..136R}
}

@ARTICLE{SIMBAD,
   author = {{Wenger}, M. and {Ochsenbein}, F. and {Egret}, D. and {Dubois}, P. and 
	{Bonnarel}, F. and {Borde}, S. and {Genova}, F. and {Jasniewicz}, G. and 
	{Lalo{\"e}}, S. and {Lesteven}, S. and {Monier}, R.},
    title = "{The SIMBAD astronomical database. The CDS reference database for astronomical objects}",
  journal = {\aaps},
   eprint = {astro-ph/0002110},
     year = 2000,
    month = apr,
   volume = 143,
    pages = {9-22},
      doi = {10.1051/aas:2000332},
   adsurl = {http://adsabs.harvard.edu/abs/2000A%26AS..143....9W}
}

@ARTICLE{heger2002,
   author = {{Heger}, A. and {Woosley}, S.~E.},
    title = "{The Nucleosynthetic Signature of Population III}",
  journal = {\apj},
   eprint = {astro-ph/0107037},
     year = 2002,
    month = mar,
   volume = 567,
    pages = {532-543},
      doi = {10.1086/338487},
   adsurl = {http://adsabs.harvard.edu/abs/2002ApJ...567..532H}
}

@ARTICLE{placco2015b,
   author = {{Placco}, V.~M. and {Beers}, T.~C. and {Ivans}, I.~I. and {Filler}, D. and 
	{Imig}, J.~A. and {Roederer}, I.~U. and {Abate}, C. and {Hansen}, T. and 
	{Cowan}, J.~J. and {Frebel}, A. and {Lawler}, J.~E. and {Schatz}, H. and 
	{Sneden}, C. and {Sobeck}, J.~S. and {Aoki}, W. and {Smith}, V.~V. and 
	{Bolte}, M.},
    title = "{Hubble Space Telescope Near-Ultraviolet Spectroscopy of Bright CEMP-s Stars}",
  journal = {\apj},
archivePrefix = "arXiv",
   eprint = {1508.05872},
 primaryClass = "astro-ph.SR",
     year = 2015,
    month = oct,
   volume = 812,
      eid = {109},
    pages = {109},
      doi = {10.1088/0004-637X/812/2/109},
   adsurl = {http://adsabs.harvard.edu/abs/2015ApJ...812..109P}
}

@ARTICLE{frebel2015,
   author = {{Frebel}, A. and {Norris}, J.~E.},
    title = "{Near-Field Cosmology with Extremely Metal-Poor Stars}",
  journal = {\araa},
archivePrefix = "arXiv",
   eprint = {1501.06921},
 primaryClass = "astro-ph.SR",
     year = 2015,
    month = aug,
   volume = 53,
    pages = {631-688},
      doi = {10.1146/annurev-astro-082214-122423},
   adsurl = {http://adsabs.harvard.edu/abs/2015ARA%26A..53..631F}
}

@ARTICLE{roederer2016,
   author = {{Roederer}, I.~U. and {Placco}, V.~M. and {Beers}, T.~C.},
    title = "{Detection of Phosphorus, Sulphur, and Zinc in the Carbon-enhanced Metal-poor Star BD+44 493}",
  journal = {\apjl},
archivePrefix = "arXiv",
   eprint = {1605.03968},
 primaryClass = "astro-ph.SR",
     year = 2016,
    month = jun,
   volume = 824,
      eid = {L19},
    pages = {L19},
      doi = {10.3847/2041-8205/824/2/L19},
   adsurl = {http://adsabs.harvard.edu/abs/2016ApJ...824L..19R}
}

@MISC{gnuplot,
      author = {Thomas Williams and Colin Kelley},
       title = {Gnuplot 5.0: an interactive plotting program},
       month = {June},
        year = {2015},
howpublished = {\href{http://www.gnuplot.info/}{http://www.gnuplot.info/}}
}

@ARTICLE{yoon2016,
   author = {{Yoon}, J. and {Beers}, T.~C. and {Placco}, V.~M. and {Rasmussen}, K.~C. and 
	{Carollo}, D. and {He}, S. and {Hansen}, T.~T. and {Roederer}, I.~U. and 
	{Zeanah}, J.},
    title = "{Observational Constraints on First-star Nucleosynthesis. I. Evidence for Multiple Progenitors of CEMP-No Stars}",
  journal = {\apj},
archivePrefix = "arXiv",
   eprint = {1607.06336},
 primaryClass = "astro-ph.SR",
     year = 2016,
    month = dec,
   volume = 833,
      eid = {20},
    pages = {20},
      doi = {10.3847/0004-637X/833/1/20},
   adsurl = {http://adsabs.harvard.edu/abs/2016ApJ...833...20Y}
}

@ARTICLE{placco2016b,
   author = {{Placco}, V.~M. and {Frebel}, A. and {Beers}, T.~C. and {Yoon}, J. and 
	{Chiti}, A. and {Heger}, A. and {Chan}, C. and {Casey}, A.~R. and 
	{Christlieb}, N.},
    title = "{Observational Constraints on First-Star Nucleosynthesis. II. Spectroscopy of an Ultra metal-poor CEMP-no Star}",
  journal = {\apj},
archivePrefix = "arXiv",
   eprint = {1609.02134},
 primaryClass = "astro-ph.SR",
     year = 2016,
    month = dec,
   volume = 833,
      eid = {21},
    pages = {21},
      doi = {10.3847/0004-637X/833/1/21},
   adsurl = {http://adsabs.harvard.edu/abs/2016ApJ...833...21P}
}

@ARTICLE{hasselquist2016,
   author = {{Hasselquist}, S. and {Shetrone}, M. and {Cunha}, K. and {Smith}, V.~V. and 
	{Holtzman}, J. and {Lawler}, J.~E. and {Allende Prieto}, C. and 
	{Beers}, T.~C. and {Chojnowski}, D. and {Fern{\'a}ndez-Trincado}, J.~G. and 
	{Garc{\'{\i}}a-Hern{\'a}ndez}, D.~A. and {Hearty}, F.~R. and 
	{Majewski}, S.~R. and {Pereira}, C.~B. and {Placco}, V.~M. and 
	{Villanova}, S. and {Zamora}, O.},
    title = "{Identification of Neodymium in the Apogee H-Band Spectra}",
  journal = {\apj},
     year = 2016,
    month = dec,
   volume = 833,
      eid = {81},
    pages = {81},
      doi = {10.3847/1538-4357/833/1/81},
   adsurl = {http://adsabs.harvard.edu/abs/2016ApJ...833...81H}
}

@INPROCEEDINGS{tody1986,
   author = {{Tody}, D.},
    title = "{The IRAF Data Reduction and Analysis System}",
booktitle = {Instrumentation in astronomy VI},
     year = 1986,
   series = {\procspie},
   volume = 627,
   editor = {{Crawford}, D.~L.},
    month = jan,
    pages = {733},
      doi = {10.1117/12.968154},
   adsurl = {http://adsabs.harvard.edu/abs/1986SPIE..627..733T}
}

@INPROCEEDINGS{tody1993,
   author = {{Tody}, D.},
    title = "{IRAF in the Nineties}",
booktitle = {Astronomical Data Analysis Software and Systems II},
     year = 1993,
   series = {Astronomical Society of the Pacific Conference Series},
   volume = 52,
   editor = {{Hanisch}, R.~J. and {Brissenden}, R.~J.~V. and {Barnes}, J.
	},
    month = jan,
    pages = {173},
   adsurl = {http://adsabs.harvard.edu/abs/1993ASPC...52..173T}
}

@book{awk,
   author = {Aho, Alfred V. and Kernighan, Brian W. and Weinberger, Peter J.},
    title = {The AWK Programming Language},
     year = {1987},
     isbn = {0-201-07981-X},
publisher = {Addison-Wesley Longman Publishing Co., Inc.},
  address = {Boston, MA, USA},
}

@INPROCEEDINGS{sed,
   author = {Lee E. Mcmahon},
    title = {SED - a non-interactive text editor},
booktitle = {UNIX Programmer’s Manual - 7th Edition, volume 2, Bell Telephone Laboratories},
     year = {1979},
publisher = {Murray Hill}
}

@ARTICLE{hansen2018,
   author = {{Hansen}, T.~T. and {Holmbeck}, E.~M. and {Beers}, T.~C. and 
	{Placco}, V.~M. and {Roederer}, I.~U. and {Frebel}, A. and {Sakari}, C.~M. and 
	{Simon}, J.~D. and {Thompson}, I.~B.},
    title = "{The R-process Alliance: First Release from the Southern Search for R-process-enhanced Stars in the Galactic Halo}",
  journal = {\apj},
archivePrefix = "arXiv",
   eprint = {1804.03114},
 primaryClass = "astro-ph.SR",
     year = 2018,
    month = may,
   volume = 858,
      eid = {92},
    pages = {92},
      doi = {10.3847/1538-4357/aabacc},
   adsurl = {http://adsabs.harvard.edu/abs/2018ApJ...858...92H}
}

@ARTICLE{roederer2018,
   author = {{Roederer}, I.~U. and {Sakari}, C.~M. and {Placco}, V.~M. and 
	{Beers}, T.~C. and {Ezzeddine}, R. and {Frebel}, A. and {Hansen}, T.~T.
	},
    title = "{The R-Process Alliance: A Comprehensive Abundance Analysis of HD222925, a Metal-poor Star with an Extreme R-process Enhancement of [Eu/H]=-0.14}",
  journal = {\apj},
archivePrefix = "arXiv",
   eprint = {1808.09469},
 primaryClass = "astro-ph.SR",
     year = 2018,
    month = oct,
   volume = 865,
      eid = {129},
    pages = {129},
      doi = {10.3847/1538-4357/aadd92},
   adsurl = {http://adsabs.harvard.edu/abs/2018ApJ...865..129R}
}

@ARTICLE{hartwig2018,
   author = {{Hartwig}, T. and {Yoshida}, N. and {Magg}, M. and {Frebel}, A. and 
	{Glover}, S.~C.~O. and {G{\'o}mez}, F.~A. and {Griffen}, B. and 
	{Ishigaki}, M.~N. and {Ji}, A.~P. and {Klessen}, R.~S. and {O'Shea}, B.~W. and 
	{Tominaga}, N.},
    title = "{Descendants of the first stars: the distinct chemical signature of second-generation stars}",
  journal = {\mnras},
archivePrefix = "arXiv",
   eprint = {1801.05044},
     year = 2018,
    month = aug,
   volume = 478,
    pages = {1795-1810},
      doi = {10.1093/mnras/sty1176},
   adsurl = {http://adsabs.harvard.edu/abs/2018MNRAS.478.1795H}
}

@book{numpy, 
      title={A guide to NumPy}, 
     author={Oliphant, Travis E}, 
     volume={1}, 
       year={2006}, 
  publisher={Trelgol Publishing USA} 
}

@ARTICLE{mendesdeoliveira2019,
   author = {{Mendes de Oliveira}, C. and {Ribeiro}, T. and {Schoenell}, W. and 
	{Kanaan}, A. and {Overzier}, R.~A. and {Molino}, A. and {Sampedro}, L. and 
	{Coelho}, P. and {Barbosa}, C.~E. and {Cortesi}, A. and {Costa-Duarte}, M.~V. and 
	{Herpich}, F.~R. and {Hernandez-Jimenez}, J.~A. and {Placco}, V.~M. and 
	{Xavier}, H.~S. and {Abramo}, L.~R. and {Saito}, R.~K. and {Chies-Santos}, A.~L. and 
	{Ederoclite}, A. and {Lopes de Oliveira}, R. and {Gon{\c c}alves}, D.~R. and 
	{Akras}, S. and {Almeida}, L.~A. and {Almeida-Fernandes}, F. and 
	{Beers}, T.~C. and {Bonatto}, C. and {Bonoli}, S. and {Cypriano}, E.~S. and 
	{Vinicius-Lima}, E. and {de Souza}, R.~S. and {Fabiano de Souza}, G. and 
	{Ferrari}, F. and {Gon{\c c}alves}, T.~S. and {Gonzalez}, A.~H. and 
	{Guti{\'e}rrez-Soto}, L.~A. and {Hartmann}, E.~A. and {Jaffe}, Y. and 
	{Kerber}, L.~O. and {Lima-Dias}, C. and {Lopes}, P.~A.~A. and 
	{Menendez-Delmestre}, K. and {Nakazono}, L.~M.~I. and {Novais}, P.~M. and 
	{Ortega-Minakata}, R.~A. and {Pereira}, E.~S. and {Perottoni}, H.~D. and 
	{Queiroz}, C. and {Reis}, R.~R.~R. and {Santos}, W.~A. and {Santos-Silva}, T. and 
	{Santucci}, R.~M. and {Barbosa}, C.~L. and {Siffert}, B.~B. and 
	{Sodr{\'e}}, L. and {Torres-Flores}, S. and {Westera}, P. and 
	{Whitten}, D.~D. and {Alcaniz}, J.~S. and {Alonso-Garc{\'{\i}}a}, J. and 
	{Alencar}, S. and {Alvarez-Candal}, A. and {Amram}, P. and {Azanha}, L. and 
	{Barb{\'a}}, R.~H. and {Bernardinelli}, P.~H. and {Borges Fernandes}, M. and 
	{Branco}, V. and {Brito-Silva}, D. and {Buzzo}, M.~L. and {Caffer}, J. and 
	{Campillay}, A. and {Cano}, Z. and {Carvano}, J.~M. and {Castejon}, M. and 
	{Cid Fernandes}, R. and {Dantas}, M.~L.~L. and {Daflon}, S. and 
	{Damke}, G. and {de la Reza}, R. and {de Melo de Azevedo}, L.~J. and 
	{De Paula}, D.~F. and {Diem}, K.~G. and {Donnerstein}, R. and 
	{Dors}, O.~L. and {Dupke}, R. and {Eikenberry}, S. and {Escudero}, C.~G. and 
	{Faifer}, F.~R. and {Far{\'{\i}}as}, H. and {Fernandes}, B. and 
	{Fernandes}, C. and {Fontes}, S. and {Galarza}, A. and {Hirata}, N.~S.~T. and 
	{Katena}, L. and {Gregorio-Hetem}, J. and {Hern{\'a}ndez-Fern{\'a}ndez}, J.~D. and 
	{Izzo}, L. and {Jaque Arancibia}, M. and {Jatenco-Pereira}, V. and 
	{Jim{\'e}nez-Teja}, Y. and {Kann}, D.~A. and {Krabbe}, A.~C. and 
	{Labayru}, C. and {Lazzaro}, D. and {Lima Neto}, G.~B. and {Lopes}, A.~R. and 
	{Magalh{\~a}es}, R. and {Makler}, M. and {de Menezes}, R. and 
	{Miralda-Escud{\'e}}, J. and {Monteiro-Oliveira}, R. and {Montero-Dorta}, A.~D. and 
	{Mu{\~n}oz-Elgueta}, N. and {Nemmen}, R.~S. and {Nilo Castell{\'o}n}, J.~L. and 
	{Oliveira}, A.~S. and {Ort{\'{\i}}z}, D. and {Pattaro}, E. and 
	{Pereira}, C.~B. and {Quint}, B. and {Riguccini}, L. and {Rocha Pinto}, H.~J. and 
	{Rodrigues}, I. and {Roig}, F. and {Rossi}, S. and {Saha}, K. and 
	{Santos}, R. and {Schnorr M{\"u}ller}, A. and {Sesto}, L.~A. and 
	{Silva}, R. and {Smith Castelli}, A.~V. and {Teixeira}, R. and 
	{Telles}, E. and {Thom de Souza}, R.~C. and {Th{\"o}ne}, C. and 
	{Trevisan}, M. and {de Ugarte Postigo}, A. and {Urrutia-Viscarra}, F. and 
	{Veiga}, C.~H. and {Vika}, M. and {Vitorelli}, A.~Z. and {Werle}, A. and 
	{Werner}, S.~V. and {Zaritsky}, D.},
    title = "{The Southern Photometric Local Universe Survey (S-PLUS): improved SEDs, morphologies, and redshifts with 12 optical filters}",
  journal = {\mnras},
archivePrefix = "arXiv",
   eprint = {1907.01567},
     year = 2019,
    month = oct,
   volume = 489,
    pages = {241-267},
      doi = {10.1093/mnras/stz1985},
   adsurl = {https://ui.adsabs.harvard.edu/abs/2019MNRAS.489..241M}
}

@ARTICLE{wolf2018,
       author = {{Wolf}, Christian and {Onken}, Christopher A. and {Luvaul}, Lance C. and
         {Schmidt}, Brian P. and {Bessell}, Michael S. and {Chang}, Seo-Won and
         {Da Costa}, Gary S. and {Mackey}, Dougal and {Martin-Jones}, Tony and
         {Murphy}, Simon J. and {Preston}, Tim and {Scalzo}, Richard A. and
         {Shao}, Li and {Smillie}, Jon and {Tisserand}, Patrick and
         {White}, Marc C. and {Yuan}, Fang},
        title = "{SkyMapper Southern Survey: First Data Release (DR1)}",
      journal = {\pasa},
         year = "2018",
        month = "Feb",
       volume = {35},
          eid = {e010},
        pages = {e010},
          doi = {10.1017/pasa.2018.5},
archivePrefix = {arXiv},
       eprint = {1801.07834},
 primaryClass = {astro-ph.IM},
       adsurl = {https://ui.adsabs.harvard.edu/abs/2018PASA...35...10W}
}

@ARTICLE{starkenburg2017,
       author = {{Starkenburg}, Else and {Martin}, Nicolas and {Youakim}, Kris and
         {Aguado}, David S. and {Allende Prieto}, Carlos and {Arentsen}, Anke and
         {Bernard}, Edouard J. and {Bonifacio}, Piercarlo and
         {Caffau}, Elisabetta and {Carlberg}, Raymond G. and
         {C{\^o}t{\'e}}, Patrick and {Fouesneau}, Morgan and
         {Fran{\c{c}}ois}, Patrick and {Franke}, Oliver and
         {Gonz{\'a}lez Hern{\'a}ndez}, Jonay I. and {Gwyn}, Stephen D.~J. and
         {Hill}, Vanessa and {Ibata}, Rodrigo A. and {Jablonka}, Pascale and
         {Longeard}, Nicolas and {McConnachie}, Alan W. and {Navarro}, Julio F. and
         {S{\'a}nchez-Janssen}, Rub{\'e}n and {Tolstoy}, Eline and
         {Venn}, Kim A.},
        title = "{The Pristine survey - I. Mining the Galaxy for the most metal-poor stars}",
      journal = {\mnras},
         year = "2017",
        month = "Nov",
       volume = {471},
       number = {3},
        pages = {2587-2604},
          doi = {10.1093/mnras/stx1068},
archivePrefix = {arXiv},
       eprint = {1705.01113},
 primaryClass = {astro-ph.GA},
       adsurl = {https://ui.adsabs.harvard.edu/abs/2017MNRAS.471.2587S}
}

@ARTICLE{bromm2009,
       author = {{Bromm}, Volker and {Yoshida}, Naoki and {Hernquist}, Lars and
         {McKee}, Christopher F.},
        title = "{The formation of the first stars and galaxies}",
      journal = {\nat},
         year = 2009,
        month = may,
       volume = {459},
       number = {7243},
        pages = {49-54},
          doi = {10.1038/nature07990},
archivePrefix = {arXiv},
       eprint = {0905.0929},
 primaryClass = {astro-ph.CO},
       adsurl = {https://ui.adsabs.harvard.edu/abs/2009Natur.459...49B}
}

@ARTICLE{placco2021,
       author = {{Placco}, Vinicius M. and {Sneden}, Christopher and {Roederer}, Ian U. and {Lawler}, James E. and {Den Hartog}, Elizabeth A. and {Hejazi}, Neda and {Maas}, Zachary and {Bernath}, Peter},
        title = "{Linemake: An Atomic and Molecular Line List Generator}",
      journal = {Research Notes of the American Astronomical Society},
         year = 2021,
        month = apr,
       volume = {5},
       number = {4},
          eid = {92},
        pages = {92},
          doi = {10.3847/2515-5172/abf651},
archivePrefix = {arXiv},
       eprint = {2104.08286},
 primaryClass = {astro-ph.IM},
       adsurl = {https://ui.adsabs.harvard.edu/abs/2021RNAAS...5...92P}
}

@ARTICLE{placco2020,
       author = {{Placco}, Vinicius M. and {Santucci}, Rafael M. and {Yuan}, Zhen and {Mardini}, Mohammad K. and {Holmbeck}, Erika M. and {Wang}, Xilu and {Surman}, Rebecca and {Hansen}, Terese T. and {Roederer}, Ian U. and {Beers}, Timothy C. and {Choplin}, Arthur and {Ji}, Alexander P. and {Ezzeddine}, Rana and {Frebel}, Anna and {Sakari}, Charli M. and {Whitten}, Devin D. and {Zepeda}, Joseph},
        title = "{The R-process Alliance: The Peculiar Chemical Abundance Pattern of RAVE J183013.5-455510}",
      journal = {\apj},
         year = 2020,
        month = jul,
       volume = {897},
       number = {1},
          eid = {78},
        pages = {78},
          doi = {10.3847/1538-4357/ab99c6},
archivePrefix = {arXiv},
       eprint = {2006.04538},
 primaryClass = {astro-ph.SR},
       adsurl = {https://ui.adsabs.harvard.edu/abs/2020ApJ...897...78P}
}

@MISC{placco2021a,
       author = {{Placco}, Vinicius M. and {Sneden}, Christopher and {Roederer}, Ian U. and {Lawler}, James E. and {Den Hartog}, Elizabeth A. and {Hejazi}, Neda and {Maas}, Zachary and {Bernath}, Peter},
        title = "{linemake: Line list generator}",
 howpublished = {Astrophysics Source Code Library, record ascl:2104.027},
         year = 2021,
        month = apr,
          eid = {ascl:2104.027},
        pages = {ascl:2104.027},
archivePrefix = {ascl},
       eprint = {2104.027},
       adsurl = {https://ui.adsabs.harvard.edu/abs/2021ascl.soft04027P}
}

@ARTICLE{bergemann2015,
       author = {{Bergemann}, Maria and {Kudritzki}, Rolf-Peter and {Gazak}, Zach and {Davies}, Ben and {Plez}, Bertrand},
        title = "{Red Supergiant Stars as Cosmic Abundance Probes. III. NLTE effects in J-band Magnesium Lines}",
      journal = {\apj},
         year = 2015,
        month = may,
       volume = {804},
       number = {2},
          eid = {113},
        pages = {113},
          doi = {10.1088/0004-637X/804/2/113},
archivePrefix = {arXiv},
       eprint = {1412.6527},
 primaryClass = {astro-ph.SR},
       adsurl = {https://ui.adsabs.harvard.edu/abs/2015ApJ...804..113B}
}

@ARTICLE{bergemann2013,
       author = {{Bergemann}, Maria and {Kudritzki}, Rolf-Peter and {W{\"u}rl}, Matthias and {Plez}, Bertrand and {Davies}, Ben and {Gazak}, Zach},
        title = "{Red Supergiant Stars as Cosmic Abundance Probes. II. NLTE Effects in J-band Silicon Lines}",
      journal = {\apj},
         year = 2013,
        month = feb,
       volume = {764},
       number = {2},
          eid = {115},
        pages = {115},
          doi = {10.1088/0004-637X/764/2/115},
archivePrefix = {arXiv},
       eprint = {1212.2649},
 primaryClass = {astro-ph.SR},
       adsurl = {https://ui.adsabs.harvard.edu/abs/2013ApJ...764..115B}
}

@ARTICLE{placco2023b,
       author = {{Placco}, Vinicius M. and {Almeida-Fernandes}, Felipe and {Holmbeck}, Erika M. and {Roederer}, Ian U. and {Mardini}, Mohammad K. and {Hayes}, Christian R. and {Venn}, Kim and {Chiboucas}, Kristin and {Deibert}, Emily and {Gamen}, Roberto and {Heo}, Jeong-Eun and {Jeong}, Miji and {Kalari}, Venu and {Martioli}, Eder and {Xu}, Siyi and {Diaz}, Ruben and {Gomez-Jimenez}, Manuel and {Henderson}, David and {Prado}, Pablo and {Quiroz}, Carlos and {Ruiz-Carmona}, Roque and {Simpson}, Chris and {Urrutia}, Cristian and {McConnachie}, Alan W. and {Pazder}, John and {Burley}, Gregory and {Ireland}, Michael and {Waller}, Fletcher and {Berg}, Trystyn A.~M. and {Robertson}, J. Gordon and {Hartman}, Zachary and {Jones}, David O. and {Labrie}, Kathleen and {Perez}, Gabriel and {Ridgway}, Susan and {Thomas-Osip}, Joanna},
        title = "{SPLUS J142445.34-254247.1: An r-process-enhanced, Actinide-boost, Extremely Metal-poor Star Observed with GHOST}",
      journal = {\apj},
         year = 2023,
        month = dec,
       volume = {959},
       number = {1},
          eid = {60},
        pages = {60},
          doi = {10.3847/1538-4357/ad077e},
archivePrefix = {arXiv},
       eprint = {2310.17024},
 primaryClass = {astro-ph.SR},
       adsurl = {https://ui.adsabs.harvard.edu/abs/2023ApJ...959...60P}
}

@ARTICLE{roederer2023,
       author = {{Roederer}, Ian U. and {Vassh}, Nicole and {Holmbeck}, Erika M. and {Mumpower}, Matthew R. and {Surman}, Rebecca and {Cowan}, John J. and {Beers}, Timothy C. and {Ezzeddine}, Rana and {Frebel}, Anna and {Hansen}, Terese T. and {Placco}, Vinicius M. and {Sakari}, Charli M.},
        title = "{Element abundance patterns in stars indicate fission of nuclei heavier than uranium}",
      journal = {Science},
         year = 2023,
        month = dec,
       volume = {382},
       number = {6675},
        pages = {1177-1180},
          doi = {10.1126/science.adf1341},
archivePrefix = {arXiv},
       eprint = {2312.06844},
 primaryClass = {astro-ph.SR},
       adsurl = {https://ui.adsabs.harvard.edu/abs/2023Sci...382.1177R}
}

@ARTICLE{astropy2018,
       author = {{\hspace{-0.15cm}Astropy Collaboration} and {Price-Whelan}, A.~M. and {Sip{\H{o}}cz}, B.~M. and {G{\"u}nther}, H.~M. and {Lim}, P.~L. and {Crawford}, S.~M. and {Conseil}, S. and {Shupe}, D.~L. and {Craig}, M.~W. and {Dencheva}, N. and {Ginsburg}, A. and {VanderPlas}, J.~T. and {Bradley}, L.~D. and {P{\'e}rez-Su{\'a}rez}, D. and {de Val-Borro}, M. and {Aldcroft}, T.~L. and {Cruz}, K.~L. and {Robitaille}, T.~P. and {Tollerud}, E.~J. and {Ardelean}, C. and {Babej}, T. and {Bach}, Y.~P. and {Bachetti}, M. and {Bakanov}, A.~V. and {Bamford}, S.~P. and {Barentsen}, G. and {Barmby}, P. and {Baumbach}, A. and {Berry}, K.~L. and {Biscani}, F. and {Boquien}, M. and {Bostroem}, K.~A. and {Bouma}, L.~G. and {Brammer}, G.~B. and {Bray}, E.~M. and {Breytenbach}, H. and {Buddelmeijer}, H. and {Burke}, D.~J. and {Calderone}, G. and {Cano Rodr{\'\i}guez}, J.~L. and {Cara}, M. and {Cardoso}, J.~V.~M. and {Cheedella}, S. and {Copin}, Y. and {Corrales}, L. and {Crichton}, D. and {D'Avella}, D. and {Deil}, C. and {Depagne}, {\'E}. and {Dietrich}, J.~P. and {Donath}, A. and {Droettboom}, M. and {Earl}, N. and {Erben}, T. and {Fabbro}, S. and {Ferreira}, L.~A. and {Finethy}, T. and {Fox}, R.~T. and {Garrison}, L.~H. and {Gibbons}, S.~L.~J. and {Goldstein}, D.~A. and {Gommers}, R. and {Greco}, J.~P. and {Greenfield}, P. and {Groener}, A.~M. and {Grollier}, F. and {Hagen}, A. and {Hirst}, P. and {Homeier}, D. and {Horton}, A.~J. and {Hosseinzadeh}, G. and {Hu}, L. and {Hunkeler}, J.~S. and {Ivezi{\'c}}, {\v{Z}}. and {Jain}, A. and {Jenness}, T. and {Kanarek}, G. and {Kendrew}, S. and {Kern}, N.~S. and {Kerzendorf}, W.~E. and {Khvalko}, A. and {King}, J. and {Kirkby}, D. and {Kulkarni}, A.~M. and {Kumar}, A. and {Lee}, A. and {Lenz}, D. and {Littlefair}, S.~P. and {Ma}, Z. and {Macleod}, D.~M. and {Mastropietro}, M. and {McCully}, C. and {Montagnac}, S. and {Morris}, B.~M. and {Mueller}, M. and {Mumford}, S.~J. and {Muna}, D. and {Murphy}, N.~A. and {Nelson}, S. and {Nguyen}, G.~H. and {Ninan}, J.~P. and {N{\"o}the}, M. and {Ogaz}, S. and {Oh}, S. and {Parejko}, J.~K. and {Parley}, N. and {Pascual}, S. and {Patil}, R. and {Patil}, A.~A. and {Plunkett}, A.~L. and {Prochaska}, J.~X. and {Rastogi}, T. and {Reddy Janga}, V. and {Sabater}, J. and {Sakurikar}, P. and {Seifert}, M. and {Sherbert}, L.~E. and {Sherwood-Taylor}, H. and {Shih}, A.~Y. and {Sick}, J. and {Silbiger}, M.~T. and {Singanamalla}, S. and {Singer}, L.~P. and {Sladen}, P.~H. and {Sooley}, K.~A. and {Sornarajah}, S. and {Streicher}, O. and {Teuben}, P. and {Thomas}, S.~W. and {Tremblay}, G.~R. and {Turner}, J.~E.~H. and {Terr{\'o}n}, V. and {van Kerkwijk}, M.~H. and {de la Vega}, A. and {Watkins}, L.~L. and {Weaver}, B.~A. and {Whitmore}, J.~B. and {Woillez}, J. and {Zabalza}, V. and {Astropy Contributors}},
        title = "{The Astropy Project: Building an Open-science Project and Status of the v2.0 Core Package}",
      journal = {\aj},
         year = 2018,
        month = sep,
       volume = {156},
       number = {3},
          eid = {123},
        pages = {123},
          doi = {10.3847/1538-3881/aabc4f},
archivePrefix = {arXiv},
       eprint = {1801.02634},
 primaryClass = {astro-ph.IM},
       adsurl = {https://ui.adsabs.harvard.edu/abs/2018AJ....156..123A}
}

@ARTICLE{astropy2013,
       author = {{\hspace{-0.15cm}Astropy Collaboration} and {Robitaille}, Thomas P. and {Tollerud}, Erik J. and {Greenfield}, Perry and {Droettboom}, Michael and {Bray}, Erik and {Aldcroft}, Tom and {Davis}, Matt and {Ginsburg}, Adam and {Price-Whelan}, Adrian M. and {Kerzendorf}, Wolfgang E. and {Conley}, Alexander and {Crighton}, Neil and {Barbary}, Kyle and {Muna}, Demitri and {Ferguson}, Henry and {Grollier}, Fr{\'e}d{\'e}ric and {Parikh}, Madhura M. and {Nair}, Prasanth H. and {Unther}, Hans M. and {Deil}, Christoph and {Woillez}, Julien and {Conseil}, Simon and {Kramer}, Roban and {Turner}, James E.~H. and {Singer}, Leo and {Fox}, Ryan and {Weaver}, Benjamin A. and {Zabalza}, Victor and {Edwards}, Zachary I. and {Azalee Bostroem}, K. and {Burke}, D.~J. and {Casey}, Andrew R. and {Crawford}, Steven M. and {Dencheva}, Nadia and {Ely}, Justin and {Jenness}, Tim and {Labrie}, Kathleen and {Lim}, Pey Lian and {Pierfederici}, Francesco and {Pontzen}, Andrew and {Ptak}, Andy and {Refsdal}, Brian and {Servillat}, Mathieu and {Streicher}, Ole},
        title = "{Astropy: A community Python package for astronomy}",
      journal = {\aap},
         year = 2013,
        month = oct,
       volume = {558},
          eid = {A33},
        pages = {A33},
          doi = {10.1051/0004-6361/201322068},
archivePrefix = {arXiv},
       eprint = {1307.6212},
 primaryClass = {astro-ph.IM},
       adsurl = {https://ui.adsabs.harvard.edu/abs/2013A&A...558A..33A}
}

@ARTICLE{aoki2015,
       author = {{Aoki}, Wako},
        title = "{Molecular Line Formation in the Extremely Metal-poor Star BD+44 493}",
      journal = {\apj},
         year = 2015,
        month = sep,
       volume = {811},
       number = {1},
          eid = {64},
        pages = {64},
          doi = {10.1088/0004-637X/811/1/64},
archivePrefix = {arXiv},
       eprint = {1507.08687},
 primaryClass = {astro-ph.SR},
       adsurl = {https://ui.adsabs.harvard.edu/abs/2015ApJ...811...64A}
}

@ARTICLE{abel2002,
       author = {{Abel}, Tom and {Bryan}, Greg L. and {Norman}, Michael L.},
        title = "{The Formation of the First Star in the Universe}",
      journal = {Science},
         year = 2002,
        month = jan,
       volume = {295},
       number = {5552},
        pages = {93-98},
          doi = {10.1126/science.295.5552.93},
archivePrefix = {arXiv},
       eprint = {astro-ph/0112088},
 primaryClass = {astro-ph},
       adsurl = {https://ui.adsabs.harvard.edu/abs/2002Sci...295...93A}
}

@ARTICLE{placco2024,
       author = {{Placco}, Vinicius M. and {Gupta}, Arvind F. and {Almeida-Fernandes}, Felipe and {Logsdon}, Sarah E. and {Rajagopal}, Jayadev and {Holmbeck}, Erika M. and {Roederer}, Ian U. and {Della Costa}, John and {Fernandez}, Pipa and {Golub}, Eli and {Higuera}, Jesus and {Patel}, Yatrik and {Ridgway}, Susan and {Schweiker}, Heidi},
        title = "{BD+44{\textdegree}493: Chemo-dynamical Analysis and Constraints on Companion Planetary Masses from WIYN/NEID Spectroscopy}",
      journal = {\apj},
         year = 2024,
        month = dec,
       volume = {977},
       number = {1},
          eid = {12},
        pages = {12},
          doi = {10.3847/1538-4357/ad8646},
archivePrefix = {arXiv},
       eprint = {2410.08943},
 primaryClass = {astro-ph.SR},
       adsurl = {https://ui.adsabs.harvard.edu/abs/2024ApJ...977...12P}
}

@ARTICLE{jeong2023,
       author = {{Jeong}, Miji and {Lee}, Young Sun and {Beers}, Timothy C. and {Placco}, Vinicius M. and {Kim}, Young Kwang and {Koo}, Jae-Rim and {Lee}, Ho-Gyu and {Yang}, Soung-Chul},
        title = "{Search for Extremely Metal-poor Stars with Gemini-N/Graces. I. Chemical-abundance Analysis}",
      journal = {\apj},
         year = 2023,
        month = may,
       volume = {948},
       number = {1},
          eid = {38},
        pages = {38},
          doi = {10.3847/1538-4357/acc58a},
archivePrefix = {arXiv},
       eprint = {2301.06236},
 primaryClass = {astro-ph.GA},
       adsurl = {https://ui.adsabs.harvard.edu/abs/2023ApJ...948...38J}
}

@ARTICLE{bonifacio2025,
       author = {{Bonifacio}, P. and {Caffau}, E. and {Fran{\c{c}}ois}, P. and {Spite}, M.},
        title = "{The most metal poor stars}",
      journal = {arXiv e-prints},
         year = 2025,
        month = apr,
          eid = {arXiv:2504.06335},
        pages = {arXiv:2504.06335},
          doi = {10.48550/arXiv.2504.06335},
archivePrefix = {arXiv},
       eprint = {2504.06335},
 primaryClass = {astro-ph.GA},
       adsurl = {https://ui.adsabs.harvard.edu/abs/2025arXiv250406335B}
}

@INPROCEEDINGS{fitzpatrick2025,
       author = {{Fitzpatrick}, Michael and {Placco}, Vinicius and {Bolton}, Adam and {Merino}, Brian and {Ridgway}, Susan and {Stanghellini}, Letizia},
        title = "{Modernizing IRAF to Support Gemini Data Reduction}",
    booktitle = {Astronomical Data Analysis Software and Systems XXXIII},
         year = 2025,
       editor = {{Jacques}, Alice and {Seaman}, Robert and {Gandilo}, Natalie and {Linder}, Tyler},
       series = {Astronomical Society of the Pacific Conference Series},
       volume = {541},
        month = oct,
        pages = {461},
          doi = {10.26624/CETF5821},
archivePrefix = {arXiv},
       eprint = {2401.01982},
 primaryClass = {astro-ph.IM},
       adsurl = {https://ui.adsabs.harvard.edu/abs/2025ASPC..541..461F}
}

@ARTICLE{placco2025,
       author = {{Placco}, Vinicius M. and {Limberg}, Guilherme and {Chiti}, Anirudh and {Prabhu}, Deepthi S. and {Ji}, Alexander P. and {Barbosa}, Fabr{\'\i}cia O. and {Cerny}, William and {Pace}, Andrew B. and {Stringfellow}, Guy S. and {Sand}, David J. and {Mart{\'\i}nez-V{\'a}zquez}, Clara E. and {Riley}, Alexander H. and {Rossi}, Silvia and {No{\"e}l}, Noelia E.~D. and {Vivas}, A. Katherina and {Medina}, Gustavo E. and {Drlica-Wagner}, Alex and {Sakowska}, Joanna D. and {Mutlu-Pakdil}, Bur{\c{c}}in and {Massana}, Pol and {Carballo-Bello}, Julio A. and {Choi}, Yumi and {Crnojevi{\'c}}, Denija and {Tan}, Chin Yi and {MAGIC Collaboration} and {Delve Collaboration}},
        title = "{The DECam MAGIC Survey: Spectroscopic Follow-up of the Most Metal-poor Stars in the Distant Milky Way Halo}",
      journal = {\apj},
         year = 2025,
        month = sep,
       volume = {991},
       number = {1},
          eid = {101},
        pages = {101},
          doi = {10.3847/1538-4357/adf846},
archivePrefix = {arXiv},
       eprint = {2506.19163},
 primaryClass = {astro-ph.GA},
       adsurl = {https://ui.adsabs.harvard.edu/abs/2025ApJ...991..101P}
}

@ARTICLE{hartwig2023,
       author = {{Hartwig}, Tilman and {Ishigaki}, Miho N. and {Kobayashi}, Chiaki and {Tominaga}, Nozomu and {Nomoto}, Ken'ichi},
        title = "{Machine Learning Detects Multiplicity of the First Stars in Stellar Archaeology Data}",
      journal = {\apj},
         year = 2023,
        month = mar,
       volume = {946},
       number = {1},
          eid = {20},
        pages = {20},
          doi = {10.3847/1538-4357/acbcc6},
archivePrefix = {arXiv},
       eprint = {2302.04366},
 primaryClass = {astro-ph.GA},
       adsurl = {https://ui.adsabs.harvard.edu/abs/2023ApJ...946...20H}
}

@Misc{mpia,
author = {{M.~Kovalev} and
{S.~Brinkmann} and {M.~Bergemann} and {MPIA IT-department}},
HOWPUBLISHED = {{NLTE MPIA web server, [Online]. Available:
{{http://nlte.mpia.de}}
Max Planck Institute for Astronomy,
Heidelberg.}},
year = {2018},
}

@ARTICLE{placco2026,
       author = {{Placco}, Vinicius M. and {Limberg}, Guilherme and {Kennedy}, Catherine R. and {Christlieb}, Norbert},
        title = "{HE0144-4657: A Carbon-Enhanced Ultra Metal-Poor Star ([Fe/H] \raisebox{-0.5ex}\textasciitilde -4.1) from the Helmi Stream Disrupted Dwarf Galaxy}",
      journal = {arXiv e-prints},
         year = 2026,
        month = jan,
          eid = {arXiv:2601.15974},
        pages = {arXiv:2601.15974},
          doi = {10.48550/arXiv.2601.15974},
archivePrefix = {arXiv},
       eprint = {2601.15974},
 primaryClass = {astro-ph.GA},
       adsurl = {https://ui.adsabs.harvard.edu/abs/2026arXiv260115974P}
}

@ARTICLE{vacca2003,
       author = {{Vacca}, William D. and {Cushing}, Michael C. and {Rayner}, John T.},
        title = "{A Method of Correcting Near-Infrared Spectra for Telluric Absorption}",
      journal = {\pasp},
         year = 2003,
        month = mar,
       volume = {115},
       number = {805},
        pages = {389-409},
          doi = {10.1086/346193},
archivePrefix = {arXiv},
       eprint = {astro-ph/0211255},
 primaryClass = {astro-ph},
       adsurl = {https://ui.adsabs.harvard.edu/abs/2003PASP..115..389V}
}

@ARTICLE{sim2014,
       author = {{Sim}, Chae Kyung and {Le}, Huynh Anh Nguyen and {Pak}, Soojong and {Lee}, Hye-In and {Kang}, Wonseok and {Chun}, Moo-Young and {Jeong}, Ueejeong and {Yuk}, In-Soo and {Kim}, Kang-Min and {Park}, Chan and {Pavel}, Michael D. and {Jaffe}, Daniel T.},
        title = "{Comprehensive data reduction package for the Immersion GRating INfrared Spectrograph: IGRINS}",
      journal = {Advances in Space Research},
         year = 2014,
        month = jun,
       volume = {53},
       number = {11},
        pages = {1647-1656},
          doi = {10.1016/j.asr.2014.02.024},
       adsurl = {https://ui.adsabs.harvard.edu/abs/2014AdSpR..53.1647S}
}

@ARTICLE{sawczynec2025,
       author = {{Sawczynec}, Erica and {Kaplan}, Kyle F. and {Mace}, Gregory N. and {Lee}, Jae-Joon and {Jaffe}, Daniel T. and {Park}, Chan and {Yuk}, In-Soo and {Chun}, Moo-Young and {Pak}, Soojong and {Hwang}, Narae and {Jeong}, Ueejeong and {Kim}, Hwihyun and {Kim}, Hyun-Jeong and {Kim}, Kang-Min and {Kim}, Sanghyuk and {Le}, Huynh Anh N. and {Lee}, Hye-In and {Lee}, Sungho and {Oh}, Heeyoung and {Oh}, Jae Sok and {Park}, Byeong-Gon and {Park}, Woojin and {Yu}, Young-Sam},
        title = "{10 Years of Archival High-resolution NIR Spectra: The Raw and Reduced IGRINS Spectral Archive (RRISA)}",
      journal = {\pasp},
         year = 2025,
        month = mar,
       volume = {137},
       number = {3},
          eid = {034505},
        pages = {034505},
          doi = {10.1088/1538-3873/adba89},
archivePrefix = {arXiv},
       eprint = {2503.05867},
 primaryClass = {astro-ph.IM},
       adsurl = {https://ui.adsabs.harvard.edu/abs/2025PASP..137c4505S}
}

@INPROCEEDINGS{oh2024,
       author = {{Oh}, Heeyoung and {Park}, Chan and {Kim}, Sanghyuk and {Kim}, Hwihyun and {Jeong}, Ueejeong and {Lee}, Hye-In and {Park}, Woojin and {Yu}, Young Sam and {Kim}, Yunjong and {Chun}, Moo-Young and {Oh}, Jae Sok and {Jang}, Jeong-Gyun and {Jang}, Bi-Ho and {Seong}, Hyeon Cheol and {Lee}, Jae-Joon and {Kim}, Hyun-Jeong and {Lee}, Sungho and {Ramos}, Francisco and {Prado}, Pablo and {Chinn}, Brian and {Arriagada}, Ignacio and {Diaz}, Ruben and {White}, John and {Tapia}, Edo and {Xu}, Siyi and {Suh}, Hyewon and {Miller}, Jennifer and {Stecher}, Hawi and {Kurz}, Emma and {Quiroz}, Carlos and {Figura}, Charlie and {Hartman}, Zachary and {Mocnik}, Teo and {Rawlings}, Mark and {Farina}, Emanuele Paolo and {Miller}, Bryan and {Stephens}, Andrew and {Oyarz{\'u}n}, Valentina and {Olivares}, Andres and {Labrie}, Kathleen and {Hirst}, Paul and {Hayward}, Thomas L. and {Brooks}, Cynthia B. and {Mace}, Gregory N. and {Lee}, Hanshin and {Good}, John M. and {Jaffe}, Daniel T. and {Kim}, Kang-Min and {Yuk}, In-Soo and {Hwang}, Narae and {Park}, Byeong-Gon},
        title = "{IGRINS-2 for Gemini Telescope: development and early performance}",
    booktitle = {Ground-based and Airborne Instrumentation for Astronomy X},
         year = 2024,
       editor = {{Bryant}, Julia J. and {Motohara}, Kentaro and {Vernet}, Jo{\"e}l. R.~D.},
       series = {Society of Photo-Optical Instrumentation Engineers (SPIE) Conference Series},
       volume = {13096},
        month = jul,
          eid = {1309608},
        pages = {1309608},
          doi = {10.1117/12.3017710},
       adsurl = {https://ui.adsabs.harvard.edu/abs/2024SPIE13096E..08O}
}

@INPROCEEDINGS{kurucz1992,
       author = {{Kurucz}, R.~L.},
        title = "{Model Atmospheres for Population Synthesis}",
    booktitle = {The Stellar Populations of Galaxies},
         year = 1992,
       editor = {{Barbuy}, Beatriz and {Renzini}, Alvio},
       series = {IAU Symposium},
       volume = {149},
        month = jan,
        pages = {225},
       adsurl = {https://ui.adsabs.harvard.edu/abs/1992IAUS..149..225K}
}

@ARTICLE{welsh2013,
       author = {{Welsh}, Barry Y. and {Montgomery}, Sharon},
        title = "{Circumstellar Gas-Disk Variability Around A-Type Stars: The Detection of Exocomets?}",
      journal = {\pasp},
         year = 2013,
        month = jul,
       volume = {125},
       number = {929},
        pages = {759},
          doi = {10.1086/671757},
       adsurl = {https://ui.adsabs.harvard.edu/abs/2013PASP..125..759W}
}

@ARTICLE{choi2025,
       author = {{Choi}, Yeon-Ho and {Jeong}, Ueejeong and {Lee}, Jae-Joon and {Kim}, Hyun-Jeong and {Oh}, Heeyoung and {Park}, Chan and {Kye}, Changwoo and {Finnerty}, Luke and {Line}, Micheal R. and {Kanumalla}, Krishna and {Sanchez}, Jorge A. and {Smith}, Peter C.~B. and {Kim}, Sanghyuk and {Lee}, Hye-In and {Park}, Woojin and {Yu}, Youngsam and {Kim}, Yunjong and {Chun}, Moo-Young and {Oh}, Jae Sok and {Lee}, Sungho and {Jang}, Jeong-Gyun and {Jang}, Bi-Ho and {Seong}, Hyeon Cheol and {Brooks}, Cynthia B. and {Mace}, Gregory N. and {Lee}, Hanshin and {Good}, John M. and {Jaffe}, Daniel T. and {Kim}, Kang-Min and {Yuk}, In-Soo and {Hwang}, Narae and {Park}, Byeong-Gon and {Kim}, Hwihyun and {Chinn}, Brian and {Ramos}, Francisco and {Prado}, Pablo and {Diaz}, Ruben and {White}, John and {Tapia}, Eduardo and {Olivares}, Andres and {Oyarzun}, Valentina and {Kurz}, Emma and {Stecher}, Hawi and {Quiroz}, Carlos and {Arriagada}, Ignacio and {Hayward}, Thomas L. and {Suh}, Hyewon and {Miller}, Jen and {Xu}, Siyi and {Farina}, Emanuele Paolo and {Figura}, Charlie and {Mocnik}, Teo and {Hartman}, Zachary and {Rawlings}, Mark and {Stephens}, Andrew and {Miller}, Bryan and {Labrie}, Kathleen and {Hirst}, Paul and {Lee}, Byeong-Cheol},
        title = "{An Early Look at the Performance of IGRINS-2 at Gemini-North with Application to the Ultrahot Jupiter, WASP-33 b}",
      journal = {\aj},
         year = 2025,
        month = oct,
       volume = {170},
       number = {4},
          eid = {238},
        pages = {238},
          doi = {10.3847/1538-3881/aded8c},
archivePrefix = {arXiv},
       eprint = {2503.12736},
 primaryClass = {astro-ph.EP},
       adsurl = {https://ui.adsabs.harvard.edu/abs/2025AJ....170..238C}
}

@Article{vanhoof2018,
AUTHOR = {Van Hoof, Peter A. M.},
TITLE = {Recent Development of the Atomic Line List},
JOURNAL = {Galaxies},
VOLUME = {6},
YEAR = {2018},
NUMBER = {2},
ARTICLE-NUMBER = {63},
URL = {https://www.mdpi.com/2075-4434/6/2/63},
ISSN = {2075-4434},
DOI = {10.3390/galaxies6020063}
}
\end{document}